\documentclass[prl,aps,twocolumn,superscriptaddress,longbibliography]{revtex4-2}
\DeclareUnicodeCharacter{2212}{\textminus}

\usepackage{lineno}
\usepackage{hyperref}
\usepackage{amsmath}

\usepackage{url}
\usepackage[caption=false,font={normalsize}]{subfig}
\usepackage{graphicx}

\newcommand{\beginsupplement}{%
    \setcounter{table}{0}
    \renewcommand{\thetable}{S\arabic{table}}
    \setcounter{figure}{0}
    \renewcommand{\thefigure}{S\arabic{figure}}
}
\usepackage{verbatim}
\usepackage{multirow}
\usepackage{amsmath, bm}
\newcommand{\beq}{\begin{eqnarray}}
\newcommand{\eeq}{\end{eqnarray}}
\usepackage{cleveref}
\usepackage{mathrsfs}
\usepackage{float}
\usepackage[usenames, dvipsnames]{color}

\usepackage{mathtools}
\usepackage{slashed}
\usepackage{physics}	
\usepackage{graphicx}   
\usepackage{epstopdf}
\usepackage{hyperref}   
\hypersetup{
pdfnewwindow=true, colorlinks=true,
linkcolor=blue, anchorcolor=blue,
citecolor=blue, filecolor=blue,
menucolor=blue, urlcolor=blue} 

\usepackage[export]{adjustbox}

\usepackage{bbold}
\usepackage{wasysym}
\usepackage{amssymb}
\makeatletter
\newcommand*{\rom}[1]{\expandafter\@slowromancap\romannumeral #1@}
\makeatother

\newcommand{\nlsm}{QNL$\sigma$M~}

\definecolor{indigo}{rgb}{0.44, 0.0, 1.0}

\definecolor{ywcolor}{rgb}{0.0, 0.5, 0.0}

\begin{document}

\title{Perturbative Kondo destruction and global phase diagram 
of heavy fermion metals}

\author{Yiming Wang}
\affiliation{Department of Physics \& Astronomy,  Extreme Quantum Materials Alliance, Smalley-Curl Institute, Rice University, Houston, Texas 77005, USA}
\author{Shouvik Sur}
\affiliation{Department of Physics \& Astronomy,  Extreme Quantum Materials Alliance, Smalley-Curl Institute, Rice University, Houston, Texas 77005, USA}
\author{Chia-Chuan Liu}
\affiliation{Department of Physics \& Astronomy,  Extreme Quantum Materials Alliance, Smalley-Curl Institute, Rice University, Houston, Texas 77005, USA}

\author{Qimiao Si}
\affiliation{Department of Physics \& Astronomy,  Extreme Quantum Materials Alliance, Smalley-Curl Institute, Rice University, Houston, Texas 77005, USA}

\begin{abstract}
Strange metals represent a foundational problem in quantum condensed matter physics, and heavy fermion systems provide a canonical setting to advance a general understanding.
The concept of a Kondo destruction quantum critical point is widely invoked to describe the competition of the Kondo effect and the local-moment magnetism.
Here, we develop a unified field-theoretic approach, analyzing this competition from a rare approach that is anchored by the magnetically ordered side.
Our analysis reveals, for the first time within a renormalization group framework, a 
quantum critical point  
across which the Kondo 
effect goes from being destroyed to dominating.
Our findings elucidate 
not only the Kondo destruction quantum criticality but also an accompanying 
global phase diagram of heavy fermion metals.
  
\end{abstract}

\maketitle

\textit{Introduction --- } 
Strange metallicity represents a fundamental subject in quantum materials research~\cite{Savitsky2025,Keimer-Moore_2017,Paschen-Si_2020,Phillips_stranger_2022,hu_quantum_2024}. 
Heavy fermion metals near an antiferromagnetic (AF) quantum critical point (QCP) represent a canonical setting for the emergence of strange metal behavior~\cite{Wir16.1,Paschen-Si_2020,kirchner_colloquium_2020}.
This phenomenon extends beyond the standard description of Landau order parameter fluctuations, namely, the spin-density-wave (SDW) order of delocalized electrons~\cite{Hertz-prb,Millis-prb}. Because the SDW model preserves quasiparticles across the bulk of the Fermi surface, it retains a Fermi-liquid character in electrical transport~\cite{Rosch99,Hlubina95}. To account for the loss of quasiparticles over the entire Fermi surface, one must instead invoke beyond-Landau critical physics in the form of a Kondo destruction QCP~\cite{Qimiao-Nature,coleman_how_2001,Senthil2004}. The defining characteristics of Kondo destruction,
such as sudden reconstruction of the Fermi surface~\cite{paschen_hall-effect_2004,shishido_drastic_2005,gegenwart_multiple_2007,friedemann_fermi-surface_2010,Custers_2012,Martelli2019}, dynamical Planckian ($\hbar \omega / k_{\rm B} T$) scaling of spin and charge responses~\cite{schroder_onset_2000,Aronson_1995,prochaska_singular_2020}
and loss of quasiparticles~\cite{chen_shot_2023,pfau_thermal_2012}, have been widely observed experimentally.

\begin{figure}[!t]
\centering
\includegraphics[scale=0.45]{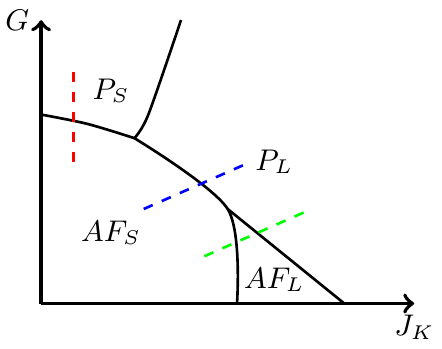}
\caption{Proposed global phase diagram for heavy fermion systems~\cite{Paschen-Si_2020}, 
where G is the degree of magnetic frustration and $J_K$ is the Kondo coupling. 
The paramagnetic ($P$) and antiferromagnetic ($AF$) phases, with Kondo screening ($L$) and destruction ($S$),
reflect the competition between the Kondo and RKKY interactions. The lines describe the associated quantum phase transitions. For details, see the main text.}

\label{fig:global}
\end{figure}

From a theoretical perspective, 
a global phase diagram---as illustrated in Fig.\,\ref{fig:global} ---has been put forward ~\cite{Paschen-Si_2020,Si-physicab-06,Qimiao-Global, Coleman-JLTP, Pixley2014,si-paschen2013quantum},
based on general considerations of the AF heavy fermion systems.
On the one hand, heavy fermion metals have traditionally been studied from a Kondo-anchored perspective~\cite{Auerbach-prl86,Millis87}.
The emphasis is the role of the AF Kondo interaction between
the local moments
and the spins of the conduction electrons,
which give rise to 
a Kondo-singlet ground state. The resulting Kondo resonances in the single-particle excitation spectrum 
participate in the formation of the Fermi surface, making it``large"~\cite{Oshikawa}; this corresponds to the $P_{\rm L}$ phase. The heavy quasiparticles on this large Fermi surface can undergo an SDW instability, leading to the AF$_{\rm L}$ phase. 
On the other hand, 
the central theme of the more recent studies has been that the 
RKKY interaction, 
which couples the local moments, can dynamically compete with the Kondo interactions described above and allow for the destruction 
of the Kondo effect, thereby leading to phases with a ``small" Fermi surface~\cite{Qimiao-Nature,coleman_how_2001,Senthil2004}. These are the paramagnetic and AF phases, P$_{\rm S}$ and AF$_{\rm S}$. The QCP between the 
P$_{\rm L}$ and AF$_{\rm S}$ phases features a complete loss of quasiparticles on the Fermi surface and strange metal behavior.

Experiments on heavy fermion metals have provided evidence for
not only the AF$_{\rm S}$ phase~\cite{paschen_hall-effect_2004,shishido_drastic_2005,gegenwart_multiple_2007,friedemann_fermi-surface_2010,Custers_2012,Martelli2019} but also the P$_{\rm S}$ phase~\cite{Hengcan-Nature}, 
where 
the Kondo effect is destroyed. This raises an important theoretical question. 
Can one approach the overall phase diagram from a magnetism-anchored  perspective? In other words, consider the RKKY interactions to have driven the local moments into an AF order, where the low-energy excitations are spin waves and the Kondo effect is destroyed, can we reach phases where the Kondo interaction succeeds in dominating over the RKKY interactions?

In this Letter, we 
take on this outstanding challenge and demonstrate such transitions. Our work not only provides the first
rigorous proof for a continuous transition between 
AF$_{\rm S}$ and P$_{\rm S}$ transitions, 
but also identify a pathway for direct transitions from either of the small-Fermi-surface phases (AF$_{\rm S}$ and P$_{\rm S}$) to the Kondo side.
We achieve these results by utilizing
a bosonic quantum nonlinear sigma model (QNL$\sigma$M) to describe the long-wavelength fluctuations of the local moments~\cite{Seiji-PRL, Chakravarty} in the Kondo lattice model.
By coupling these bosonic fluctuations to itinerant fermions via the Kondo coupling, we construct a controlled renormalization group (RG) framework that treats both degrees of freedom on an equal footing, 
reflecting the spirit 
of 
the studies in
metallic  
systems
~\cite{Polchinski1994, Nayak1994, Abanov2004,SSLee,Metlitski1,Metlitski2,Mross2010,Sur2014,Sur2015,Sur2016,SSLee2}. 
This unified treatment distinguishes the present analysis from earlier approaches to Kondo lattices~\cite{Seiji-PRL},
enabling us to capture, within a single RG framework, the emergence of the Kondo-destruction QCP and the global connectivity
in the phases of the global phase diagram.

\textit{The Model ---} The $d$-dimensional Kondo lattice model is represented in a mixed basis, 
 \begin{align}\label{konla}
H=\sum_{\vec k,\sigma, a}E_{\vec k}\psi^{\dagger}_{\vec k,\sigma,a}\psi_{\vec k,\sigma,a}
+ \sum_{ij}I_{ij}\vec{S}_i\cdot\vec{S}_j
+J_K\sum_{i,a}\vec{S}_{i}\cdot\vec{s}_{i,a},
\end{align}
where  $\vec k$ ($i,j$) represents momentum (position space lattice coordinates), $\vec{S}_i$ and $\vec{s}_{i,a}=\sum_{\alpha\beta}\psi^{\dagger}_{i,\alpha,a}\vec{\sigma}_{\alpha\beta}\psi_{i,\beta,a}$ are the spin of the local moments and spin-density of the conduction electrons on channel $a$ with $a=1,...,M$, respectively, on the $i$-th site,  $\psi_{k,\sigma,a}$ is the electron destruction operator acting at momentum $\vec k$ and on spin-index $\sigma$ and channel-index $a$, and $E_{\vec{k}}$ is the band-dispersion of the conduction electrons. 
We have introduced the multiplicity of channels, motivated by recent developments in Bose-Fermi Kondo models~\cite{hu2207kondo}, with $M$ fixed to $2S$.
$I_{ij}$ ($J_K$) represents the strength of the RKKY  (Kondo) interaction between local moments (local moments and electronic spin-density.
Here, we assume these couplings to be positive, i.e. antiferromagnetic.
In order to minimally incorporate the effects of frustration, we consider  only nearest and next-nearest neighbor couplings $I_{1}$ and $I_{2}$, with the degree of frustration being controlled by the ratio $I_2/I_1$.

In the limit $J_K = 0$, the  local-moment sector is described by the Heisenberg model with nearest and next-nearest neighbor interactions. 
As an example, we recall that on the square lattice in both $I_1 \gg I_2$ and $I_1 \ll I_2$ limits antiferromagnetically ordered states are obtained with the ordering wavevector $\vec Q = (\pi, \pi)$ and $(0, \pi)$, respectively. 
Classically, a phase transition $I_2/I_1 = 1/2$ separates the two ordered states~\cite{Chandra1988}.
Quantum fluctuations  opens up an extended parameter window about this point where quantum paramagnetic states have been observed in large scale numerical simulations~\cite{gong2014,richter2010spin}.
{Accordingly, we can define the frustration parameter for the square lattice model as $G =2/(2I_2/I_1 + I_1/2 I_2)$} such that $G$ is maximized at $I_2/I_1 = 1/2$.

 In this work, for concreteness, we consider the square lattice and start from the 
 AF region with $I_2/I_1 < 1/2$ with $\vec Q = (\pi, \pi)$.
For simplicity, we assume the conduction electron's own Fermi surface to not intersect the magnetic zone boundary (i.e., its
$K_F$ does not reach the AF ordering wavevector $|\vec Q|$). 
In order to locate at and in the vicinity of the 
AF
phase we impose $J_K$ to be small compared to both $I$ and $W$ ($I \ll W$), where $W$ is the bandwidth of the electron spectrum.
Since we are interested in the low energy behavior of the Kondo lattice model in Eq.~\eqref{konla}, we will obtain an effective action that governs the long-wavelength fluctuations in the system.  
This is achieved through a two step process.
First, we generalize the spin of the local moment to $S$ and decompose $\vec{S}_i$ into a three component vector field  $\vec{n}\left(\vec{x},\tau\right)$ which tracks the fluctuations in the staggered component of the local moments, and a canting field $\vec{L}\left(\vec{x},\tau\right)$ that describes the uniform component of local-moment fluctuations~\cite{Chakravarty,Haldane-PLA, Auerbach},
\begin{align}
 \vec{S}_i/S\rightarrow e^{i\vec{Q}\cdot\vec{x}}\vec{n}\left(\vec{x},\tau\right) \sqrt{1-\left(2a^d\vec{L}\left(\vec{x},\tau\right)\right)^2}+2a^d\vec{L}\left(\vec{x},\tau\right),
\end{align}
where  $\vec{x}$ labels the position and $a$ is the lattice constant. 
In order to preserve the norm  $|\vec S_i| = S$ we require  $\vert \vec{n}\left(\vec{x},\tau\right)\vert=1$ and $\vec{n}\left(\vec{x},\tau\right)\cdot\vec{L}\left(\vec{x},\tau\right)=0$.
Integrating out $\vec L$ induces the \nlsm in the pure local-moment sector~\cite{Seiji-PRL}, and the microscopic Kondo coupling, $J_{\text K}$, generates a coupling between  $\vec n$ and the conduction electrons. 
We note that at this stage the effective theory describes the \nlsm coupled to conduction electrons~\cite{sm}; consequently, it exemplifies a constrained field theory, in analogy to the QNL$\sigma$M. 

In the second step of the derivation, we parameterize the constrained field $\vec n$ as 
$\vec{n}=\left(\vec{\pi},\sigma\right)$ with $\sigma=\sqrt{1-\vec{\pi}^2}$.
The vector field  $\vec{\pi}=\left(\pi_1,\pi_2\right)$ represents the magnon modes that are generated by fluctuations about the staggered ordering of the local moments.
Here, we will assume these fluctuations are sufficiently weak such that $\sqrt{1-\vec{\pi}^2} \cong 1-\frac{1}{2}\left(\pi_{1}^2+\pi_{2}^2\right)+\cdots$.
Thus, $\vec \pi$ captures the long wavelength fluctuations of $\vec n$ about the ordering direction, $\vec n_{\text{ordered}} = (0, 0, 1)$.
It is convenient to scale the magnon field $\vec{\pi} \to \vec \phi \equiv \vec \pi / \sqrt{g}$ such that the propagator of $\vec \phi$ takes the conventional form. 
Therefore, the long wavelength dynamics of the Kondo lattice model is expressed in terms of $\vec \phi$ and the conduction electrons, $\psi$,
\begin{align} \label{model}
S =&S_c+S_n+S_K\\
S_c = &\sum_{\sigma, a}\int d^dKd\Omega\psi^{\dagger}_{\vec{K}\Omega,\sigma,a}\left(-i\Omega+E_{\vec{K}} - \mu\right)\psi_{\vec{K}\Omega,\sigma,a}\nonumber\\
S_{n}=&\frac{1}{2}\int d\tau d^{d}x\{[(\partial_{\tau}\vec{\phi})^{2}+c^{2}(\vec{\nabla}\vec{\phi})^{2}] \nonumber \\
&+g[(\vec{\phi}\cdot\partial_{\tau}\vec{\phi})^{2}+c^{2}(\vec{\phi}\cdot\vec{\nabla}\vec{\phi})^{2}]\}\nonumber,\\ 
S_{K}=& \sum_a \int d\tau d^dx\{i\lambda\epsilon_{\alpha\beta}\partial_{\tau}\phi_{\alpha}s_{a}^{\beta}+i\lambda_{z}\epsilon_{\alpha\beta}\phi_{\alpha}\partial_{\tau}\phi_{\beta}s_{a}^{z}\} \nonumber.
\end{align}
where $\mu$ is the chemical potential, $g=4 a^{d}d(I_{1}+2I_{2})$, $c=2aS\sqrt{d}\sqrt{(I_1+2I_{2})(I_1-2I_2)}$, and $\epsilon_{\alpha\beta}$ is the totally anti-symmetric two-component tensor.
The temporal derivatives in the electron-boson interaction vertices originate from the spin Berry phase term in the QNL$\sigma$M, and their signs are fixed by the relative sign between the Berry phase term and the Kondo coupling.
Since the corresponding couplings, $\lambda$ and $\lambda_z$, share a common origin in the Kondo vertex [$J_{\text K}$ in Eq.~\eqref{konla}], they obtain similar forms, $\{\lambda, \lambda_{z}\} = \{\sqrt{g}, g \}  J_{\text K} Sa/4d$, at the ultra-violet fixed point. 
Importantly, rotational symmetry about the ordered moment imposes a strict relation between these couplings,
 \begin{align}
\lambda_z=\sqrt{g}\,\lambda.
 \end{align}
As shown below, this constraint is consistent with the renormalization group (RG) flow, allowing us to work with a single effective coupling $\lambda$ in the following analysis.

In contrast to the well-known results on the low-energy dynamics of the pure \nlsm model~\cite{Chakravarty}, the fate of the system when the \nlsm is coupled to itinerant electrons remains unclear. 
While the presence of the derivatives in the boson-fermion vertices makes these interactions weaker than more conventional models of metallic quantum criticality at low energies~\cite{SSLee2}, strong frustration and microscopic Kondo interaction introduces  a non-trivial interplay between the fluctuations of the local moments and the conduction electrons. 
Indeed, previous works~\cite{Seiji-PRL, Yamamoto-PRB}
have   indicated the capacity of the fluctuations in the  Kondo lattice model to fundamentally alter the infrared dynamics due to kinematic constraints imposed by the Fermi surface.
(We also note on the importance of not to integrate out the fermions~\cite{Ong-PRL} from the Kondo lattice.)
These works, however, do not adequately address the physics at the edge of the 
AF phase which is foundational to the phase diagram of the Kondo lattice model (cf. Fig.~\ref{fig:global}).
In the scaling analysis to follow, we  address this open question through a systematic RG analysis of the effective action in Eq.~\ref{model}.

\begin{figure}[h!]
\centering
\includegraphics[scale=0.3]{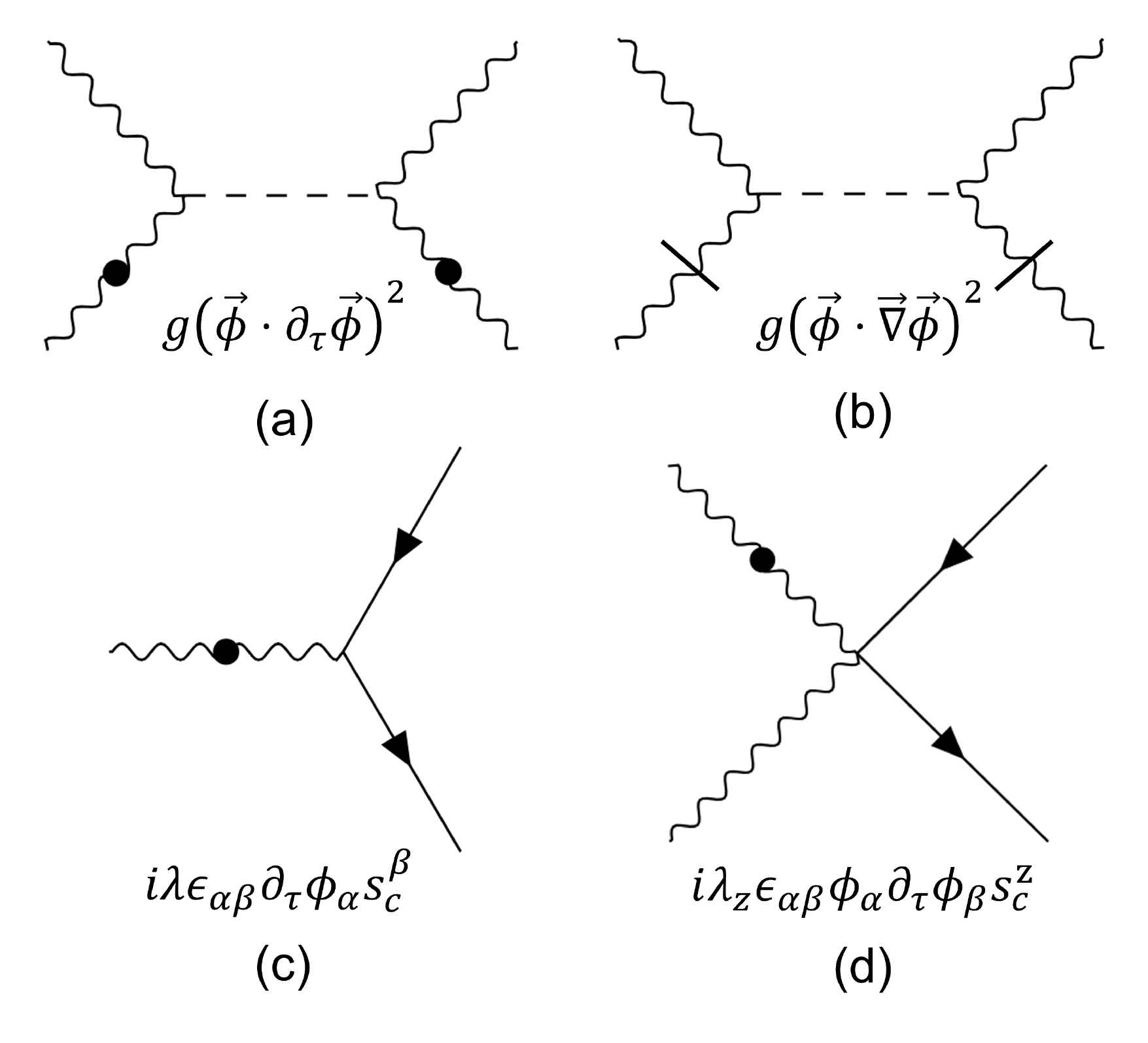}
\caption{Feynman rules of the action (\ref{model}) after rescaling $\vec{\pi}\rightarrow\sqrt{g}\vec{\phi}$, where $\alpha=x,y$. The solid arrow line and the wavy line are the propagators of the fermionic and bosonic field respectively. The slash and dot on the bosonic propagators denote the space and time derivative respectively. Expansion to higher orders contains vertices with more boson legs, where the leading order of them still contributes to (c) and (d) through vertex corrections. We summarize the vertex corrections from higher order vertices through Supplementary Materials~\cite{sm}.
}
\label{fig:feyrule}
\end{figure}

\textit{Renormalization group analysis ---} The propagators of the boson and fermion fields are, respectively, 
\begin{align}
D\left(\vec{q},i\omega\right)=\frac{1}{\omega^2+c^{2}\vec{q}^{\,2}}, \qquad G_{\psi}(\vec{K},i\Omega)=\frac{1}{i\Omega-v k_{\perp}}
\end{align}
where $v$ is the Fermi velocity and $k_{\perp}=|\vec{K}|-K_{F}$ is the distance from the Fermi surface in momentum space. 
A combination of Wilson's bosonic scaling and Shankar's fermionic scaling for $z=1$~\cite{Yamamoto-PRB,huajia2014}, determines the following energy scaling dimensions
\begin{align}
    &[\omega]=[k_{\perp}]=[\vec{q}]=1,\\
    &[\vec{\phi}(\vec{q},i\omega)]=-\frac{d+3}{2},\,[\psi(\vec{K},i\Omega)]=-\frac{3}{2}
\end{align}
at the tree level. 
Within this scheme $[g]=-\epsilon$, and $[\lambda]=[g]/2=-\epsilon/2$, where $\epsilon=d-1$. 
Therefore for $d>1$, the
AF phase is perturbatively stable against the electron-boson couplings. 
Quantum corrections due to sufficiently strong couplings, however, can destabilize the 
AF phase towards itinerant quantum paramagnetic states. 
In order to track this competition, we perform an RG analysis at one-loop order.  
For simplicity, we will focus on the limit $v/c \ll 1$ which suppresses the generation of symmetry-allowed vertices beyond those included in the effective action, Eq.~\eqref{model}.

\begin{figure}[!t]
\centering
\includegraphics[scale=0.3]{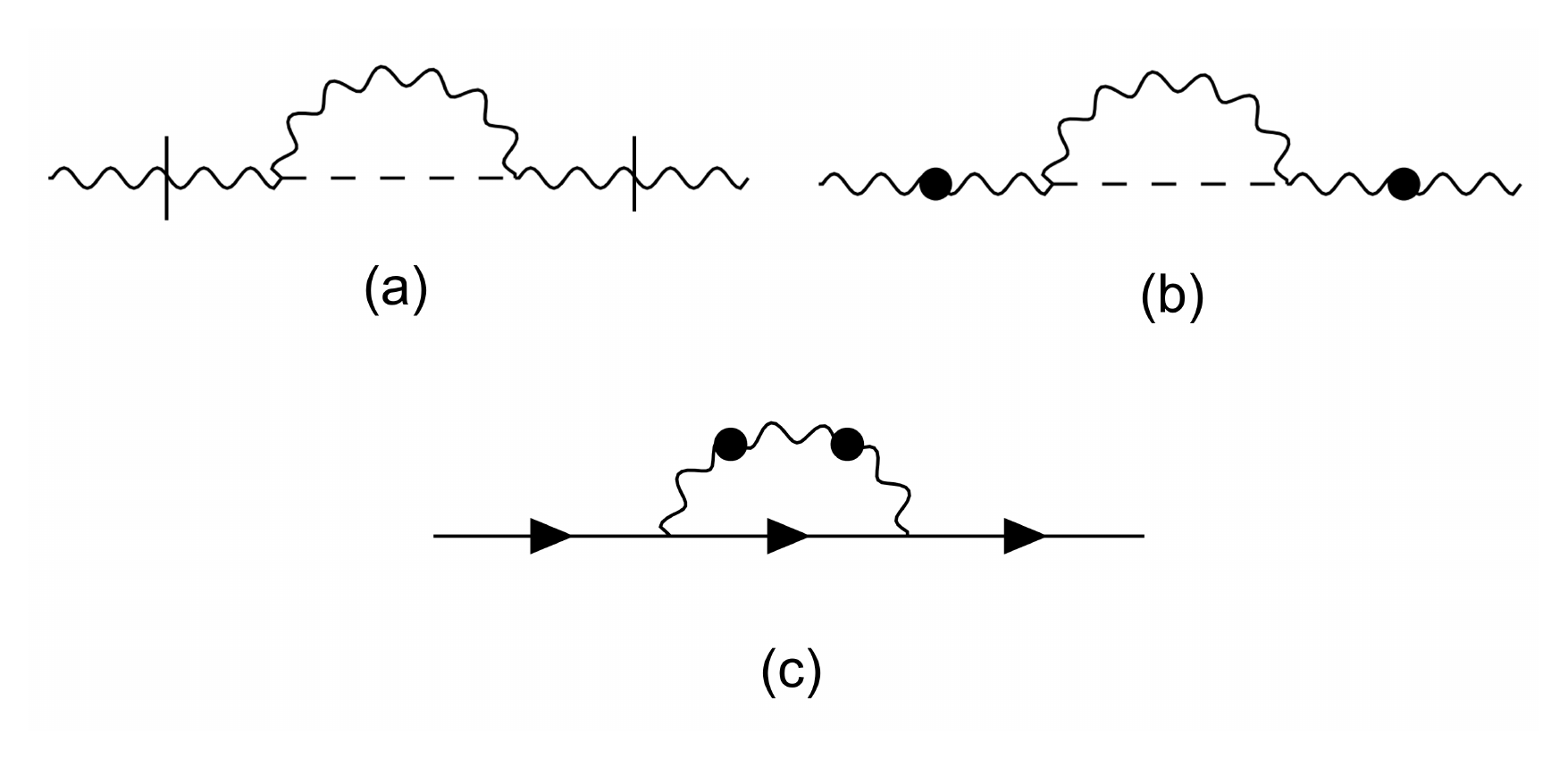}
\caption{Singular one-loop diagrams for (a,b) bosonic propagator  (c) for fermion propagator 
for transverse Kondo coupling. \label{fig:selfenergy-MT}}
\end{figure}

\begin{figure}[!t]
\centering
\includegraphics[scale=.28]{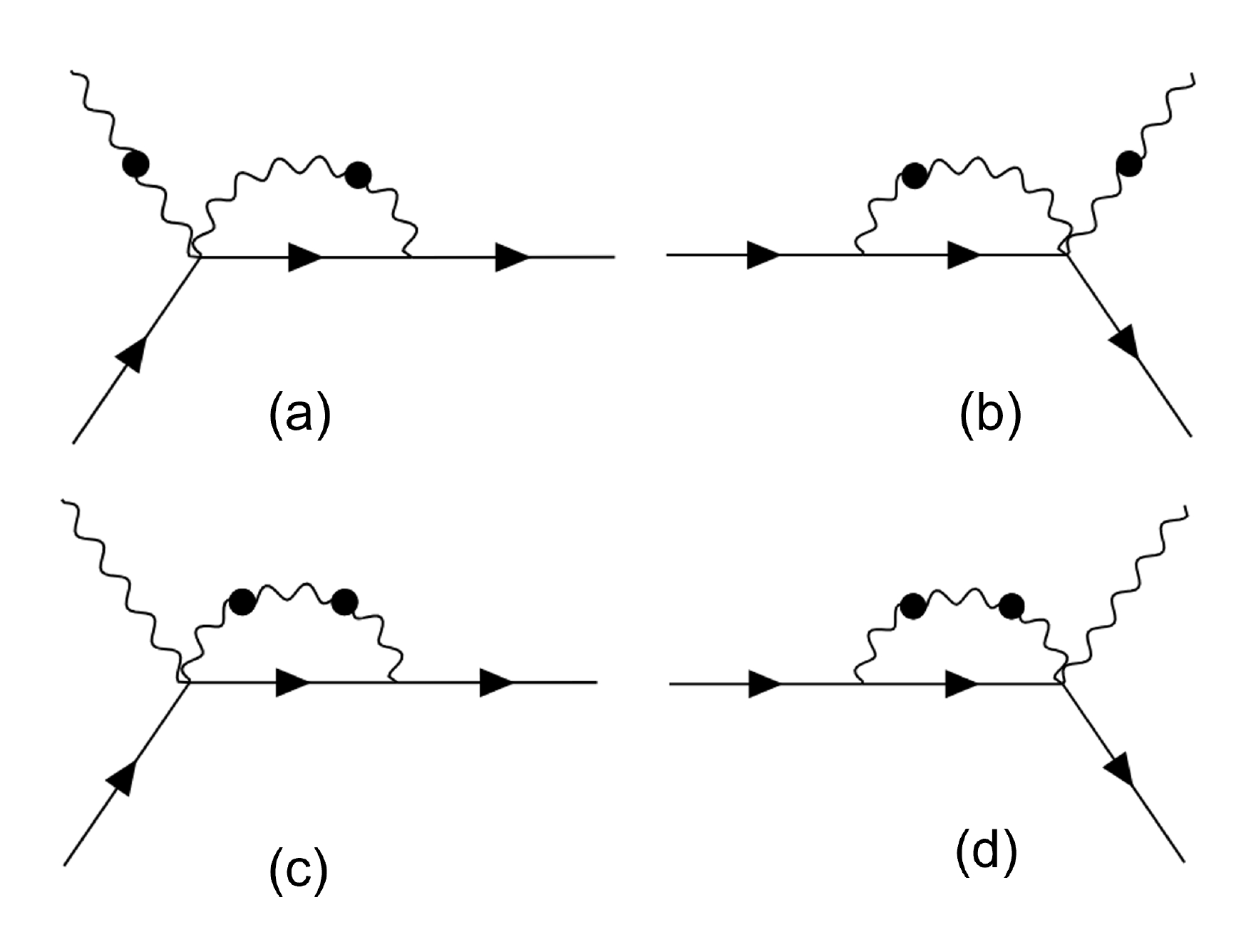}
\caption{Vertex corrections at one loop level. Only the diagrams shown here contribute non-trivially to the beta function; other topologically possible one-loop diagrams vanish (see Supplementary Material, Fig. S4 for the complete classification)}
\label{fig:vc}
\end{figure}

The log-divergent Feynman diagrams that contribute non-trivially to the $\beta$ functions of $g$ and $\lambda$ at one-loop order form a subset of all possible one-loop processes, and they are listed  in Figs.~\ref{fig:selfenergy-MT} and \ref{fig:vc}. 
The details of their evaluation, along with a comprehensive list of all one-loop diagrams,  are given in the Supplemental Material~\cite{sm}. 
We use dimensional regularization within the minimal subtraction scheme to obtain the $\beta$ functions,
\begin{align}
\beta(g)&=-\epsilon g+g^{2},\label{rgeq1}\\
\beta(\lambda)&=-\tfrac{1}{2}(\epsilon+g)\lambda-\lambda^{3}
+\tfrac{3\sqrt{2}}{2}\tfrac{1+2r/3}{1+r}\sqrt{g}\,\lambda^{2}.\label{rgeq2}
\end{align}
where $\beta(g)\equiv dg/d\ell$ and $\beta(\lambda)\equiv d\lambda/d\ell$ with $\ell$ being the logarithmic length scale, $r=v/c$ is the ratio between the Fermi and spin velocities, and we have rescaled $g\!\to\!2\pi c g$ and $\lambda\!\to\!\lambda/\sqrt{2A}$ with $A=c/(2\pi (c+v)^{2})$. 
We note that $v$ and $c$ do not flow at one-loop order. 
Eq.~(\ref{rgeq1}) reproduces the well-known flow of the quantum nonlinear sigma model: the linear term represents the scaling dimension of the spin stiffness, while the $g^2$
 term reflects the strengthening of magnon–magnon interactions at low energies, driving the magnetic quantum phase transition. The flow of the Kondo coupling in Eq.~(\ref{rgeq2}) contains three distinct  contributions. 
The term proportional to $-\lambda$ gives the tree-level scaling dimension inherited from the fermion and boson fields.  
The $-\lambda^{3}$ term arises from the one-loop fermion self-energy diagram in Fig.~\ref{fig:selfenergy-MT}(c).  
Finally, the $\sqrt{g}\lambda^{2}$ term originates from the vertex corrections in Fig.~\ref{fig:vc}; this contribution is unique to the dynamical coupling between fermions and quantum spin fluctuations, and it is the key ingredient responsible for generating the non-trivial interacting fixed points discussed below.

In deriving the flow equations above, we have used the rotational-symmetry constraint 
$\lambda_{z}=\sqrt{g}\,\lambda$. 
A direct evaluation of the one-loop vertex corrections shows that 
$\beta(\lambda_{z})=\beta(\sqrt{g}\lambda)$~\cite{sm}, confirming that this relation is preserved by the RG flow. 
It is therefore consistent to impose $\lambda_{z}=\sqrt{g}\,\lambda$ throughout the analysis and work with a single effective Kondo coupling $\lambda$.

\begin{figure}[!t]
\centering
\includegraphics[scale=.8]{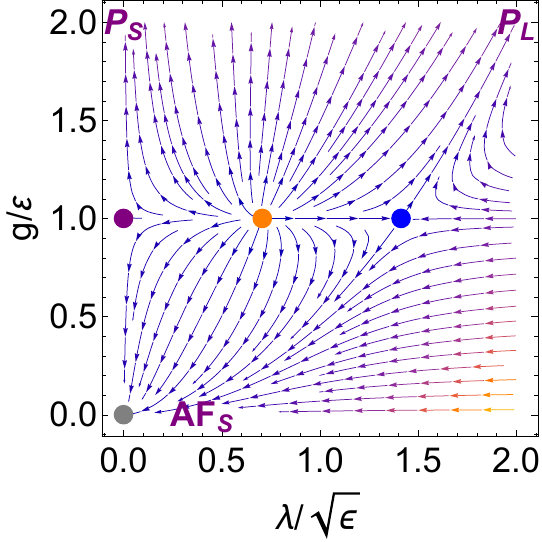}
\caption{RG flow of $g$ and $\lambda$ for $v/c<(\sqrt{2}-1)/2$, where the gray dot represents the 
AF fixed point, and the purple dot denotes the QCP of QNL$\sigma$M at $g^*=\epsilon$. The orange and blue dots denote for multicritical point and Kondo destruction critical point. }
\label{fig:RGflow}
\end{figure}

With the RG equations in hand, we now analyze the fixed-point structure that governs the competing influences of magnetic fluctuations and the Kondo interaction.  
The flow diagram in Fig.~\ref{fig:RGflow} summarizes our results.
\textit{(i) Fixed points at $\lambda=0$.}  
In the absence of Kondo coupling, $\lambda=0$, the RG equations reduce to those of the quantum nonlinear sigma model.  
We recover a stable antiferromagnetic fixed point at $(g^{*}=0,\lambda^{*}=0)$ and the conventional magnetic 
QCP at $(g^{*}=\epsilon,\lambda^{*}=0)$, which controls the transition between the antiferromagnet and a paramagnetic phase with a small Fermi surface ($P_{S}$).  
These two fixed points reproduce the standard magnetic criticality of the QNL$\sigma$M.
\textit{(ii) Interacting fixed points at $\lambda>0$.}  
Once the antiferromagnetic Kondo coupling is turned on, the vertex correction term in Eq.~(\ref{rgeq2}) qualitatively changes the flow.  
Along the magnetic critical line $g^{*}=\epsilon$, two non-trivial interacting fixed points appear at
\begin{align}
\lambda^{*}= f_{\pm}(r) \sqrt{\epsilon},
\end{align}
where
$
    f_{\pm}(r)=\sqrt{\frac{1}{1+r}+\frac{1}{4(1+r)^{2}}
    \pm\sqrt{\left[\frac{1}{1+r}+\frac{1}{4(1+r)^{2}}\right]^{2}-1}}.
$
The larger root, $f_{+}(r)$, corresponds to the Kondo-destruction
QCP.
This fixed point is unstable and separates RG flows toward two distinct ground states.  
Flows toward larger $\lambda$ indicate that the Kondo effect becomes dominant, leading to a heavy-Fermi liquid with a large Fermi surface ($P_{L}$).  
Flows toward smaller $\lambda$ drive the system back to magnetic order.  
The instability of this fixed point embodies the breakdown of Kondo screening at criticality.
The smaller root, $f_{-}(r)$, corresponds to a quantum multicritical point.  
This fixed point is unstable along all directions in the $g-\lambda$ plane  and forms the nexus connecting all three phases: the antiferromagnet ($AF_{S}$), the paramagnet with a small Fermi surface ($P_{S}$), and the heavy-fermion liquid with a large Fermi surface ($P_{L}$).  
Its appearance in the perturbative RG mirrors the structure long anticipated in the heavy-fermion global phase diagram (c.f. Fig.~\ref{fig:global}).

\textit{Critical exponents and loss of quasiparticles ---}  At the interacting fixed points identified above, the fermionic and bosonic propagators take the asymptotic forms
\begin{align}
D(\omega,q) &\sim \frac{1}{(\omega^2 + c^2 q^2)^{1-\gamma_\phi}},\\
G_\psi(\Omega,k_\perp) &\sim \frac{1}{(i\Omega - v k_\perp)^{1-2\gamma_\psi}},
\end{align}
with the details of the derivations of the boson ($\gamma_\phi$) and fermion ($\gamma_\psi$) anomalous dimensions presented in the SM~\cite{sm}.
{Since the anomalous dimensions at the \nlsm QCP is well-known [$(\gamma_\phi, \gamma_\psi) = (\epsilon/2, 0)$, indicating critical magnon fluctuations against a Fermi liquid background], here, we focus on the two new fixed points.}
At the Kondo-destruction QCP  ($\lambda^* =  f_+ \sqrt{\epsilon}$), 
\begin{equation}
\gamma_\phi = \frac{\epsilon}{2} + O(\epsilon^2), \qquad 
\gamma_\psi = f_+(r)^2\frac{\epsilon}{2} + O(\epsilon^2),
\end{equation}
while at the multicritical point at $\lambda^* =  f_- \sqrt{\epsilon}$,
\begin{equation}
\gamma_\phi = \frac{\epsilon}{2} + O(\epsilon^2), \qquad
\gamma_\psi = f_-(r)^2\frac{\epsilon}{2} + O(\epsilon^2)
\end{equation}
We note that the non-zero fermion anomalous dimension signals the breakdown of electron quasiparticles, whereas the non-zero boson exponent reflects critical magnon fluctuations.

\textit{Discussions.~~} Several remarks are in order. First, these non-trivial fixed points (Kondo destruction QCP and multicritical point) emerge only for an antiferromagnetic Kondo interaction ($\lambda > 0$) through a delicate interplay between the Berry phase term and the Kondo coupling that is reflected in a 
competition between Kondo singlet formation and magnetic ordering.
For a ferromagnetic coupling ($\lambda < 0$), the Kondo-interaction is irrelevant, and no such critical points are found.

Second, the controlled renormalizability of our field theory restricts us to the regime where 
$r=v/c \ll 1$.
While a nonzero $r$ generates new vertices through quantum fluctuations, these vertices remain parametrically suppressed through the course of the RG flow as long as $r \ll 1$.
We note that this regime is consistent with the large-$S$ limit of the underlying spin model, which results in $r \sim 1/S$ as the spin-wave velocity $c$ grows with $S$ while the electronic velocity $v$ remains fixed.   
A systematic treatment of the general case with an arbitrary $r$ is left for future work.

Third, the experiment on CePdAl, a heavy-fermion metal with a distorted Kagome lattice~\cite{Hengcan-Nature}, provided the first experimental evidence for a paramagnetic small-Fermi-surface ($P_S$) phase and a multicritical point connecting the $AF_S$, $P_S$, and $P_L$ regimes. 
The strong spin frustration in CePdAl suppresses conventional N\'eel order and enhances quantum fluctuations, creating favorable conditions for Kondo destruction to occur on the antiferromagnetic side of the phase diagram. Our RG analysis, formulated from the antiferromagnetic limit, captures this interplay between Kondo screening and frustrated magnetism, yielding a Kondo-destruction QCP and reproducing the global phase diagram supported by experiments. 
This provides a unified field-theoretic perspective.

\textit{Conclusion and Outlook.}
We have developed a renormalization-group framework for the Kondo lattice model, formulated from the antiferromagnetic side, which identifies a Kondo-destruction quantum critical point and  
captures the global phase diagram encompassing the $AF_S$, $P_S$, and $P_L$ phases (c.f. Fig.~\ref{fig:global}). 
This work provides the first  
comprehensive understanding of the proposed global phase diagram.
Beyond the context of heavy-fermion metals such as CePdAl, the theoretical framework established here can be applied to designer quantum simulators and flat-band systems where strong correlations and magnetic frustration coexist. Such platforms offer a promising route to experimentally explore quantum criticality and Kondo
destruction in a highly tunable setting.

\acknowledgements 
We thank R. Pelcovits for a helpful correspondence, and S. Paschen and F. Steglich for discussion and motivation.
This work has been supported in part by
the NSF Grant No.\ DMR-2220603, 
and by the Robert A. Welch Foundation Grant No.\ C-1411 
and 
the Vannevar Bush Faculty Fellowship ONR-VB N00014-23-1-2870. 
Q.S. acknowledges the hospitality of the Aspen Center for Physics, which is supported by NSF grant No. PHY-2210452.

\bibliography{criticality}
\bibliographystyle{apsrev4-2}

\clearpage
\onecolumngrid

\setcounter{secnumdepth}{3}

\onecolumngrid
\newpage
\beginsupplement
\section*{Supplemental Materials}

\numberwithin{equation}{section} 

\tableofcontents

\onecolumngrid

\section{Low-energy effective action}

To derive the effective field theory of the Kondo lattice model in the antiferromagnetic regime, we employ the coherent-state path integral formalism for both local-moment spins and conduction electrons. This framework allows us to access the long-wavelength dynamics of the spin degrees of freedom and their coupling to itinerant fermions, leading to the quantum non-linear sigma model (QNL$\sigma$M) with Kondo interaction.

The Euclidean action consists of a spin Berry-phase term and the Hamiltonian contributions:
\begin{align}
S &= S_{B} + \int d\tau\, H(\tau),\\
S_{B} &= iS \sum_{i} \int_{0}^{\beta} d\tau \int_{0}^{1} du \,
\vec{\Omega}_{i} \cdot 
\left( \frac{\partial \vec{\Omega}_{i}}{\partial u} 
\times 
\frac{\partial \vec{\Omega}_{i}}{\partial \tau} \right).
\end{align}
The Hamiltonian is composed of three parts: $H(\tau) = H_{n}(\tau) + H_{c}(\tau) + H_{K}(\tau)$. The local moment interactions are given by
\begin{align}
H_{n}(\tau) &= S^{2} I_{1} \sum_{\langle i, j\rangle} 
\vec{\Omega}_{i} \cdot \vec{\Omega}_{j}
+ S^{2} I_{2} \sum_{\langle\langle i,j \rangle\rangle}
\vec{\Omega}_{i} \cdot \vec{\Omega}_{j}.
\end{align}
The conduction electron hopping and Kondo coupling are
\begin{align}
H_{c}(\tau) &= -t \sum_{\langle i, j\rangle,\sigma,a} 
\left( c^{\dagger}_{i,\sigma,a} c_{j,\sigma,a} + \text{h.c.} \right)
- \mu \sum_{i,\sigma,a} c^{\dagger}_{i,\sigma,a} c_{i,\sigma,a},\\[4pt]
H_{K}(\tau) &= J_{K} S \sum_{i,a} 
\vec{\Omega}_{i} \cdot \vec{s}_{c,a,i},
\end{align}
where $\vec{s}_{c,a,i} = c^{\dagger}_{i,\mu} \vec{\tau}_{\mu\nu} c_{i,\nu}$ is the electron spin density at site $i$,channel $a$, with $a=1,\cdots,M$.

\vspace{0.5em}
\noindent\textbf{Antiferromagnetic Decomposition.}
We separate the spin configuration at site $i$ into a slowly varying staggered component $\vec{n}(\vec{x}_i,\tau)$ and a small uniform component $\vec{L}(\vec{x}_i,\tau)$:
\begin{align}
 \vec{\Omega}_i \approx \eta_i \vec{n}(\vec{x}_i,\tau) + 2a_0^d \vec{L}(\vec{x}_i,\tau),
\end{align}
where $\eta_i = e^{i\vec{Q}\cdot\vec{x}_i}$ is the staggering factor. The fields satisfy $\vec{n}^{2} = 1$ and $\vec{n}\cdot\vec{L} = 0$.

\vspace{0.5em}
\noindent\textbf{Continuum Expansion.}
We expand the action in the continuum limit $a_0 \to 0$:

1. \textit{Berry Phase:} The Berry phase decomposes into two distinct contributions:
\begin{align}
S_{B} \approx S_{\text{WZ}}[\vec{n}] - i \frac{S}{a_0^d} \int d\tau d^dx \, \vec{L} \cdot (\vec{n} \times \partial_\tau \vec{n}).
\end{align}
The first term is the topological Wess-Zumino term, which we neglect as the contributions from the two sublattices cancel for a bipartite antiferromagnet. The second term is the dynamical coupling between the Néel order and the uniform magnetization. The negative sign is consistent with the standard Euclidean path-integral convention found in established literature~\cite{subir1990, fradkin2021, Sachdev_2011}. 

2. \textit{Heisenberg Interaction:} The exchange terms generate a stiffness for the Néel field and a penalty for uniform magnetization:
\begin{align}
H_{n} \rightarrow \int d^dx \left[ \frac{\rho_s}{2} (\nabla \vec{n})^2 + \frac{1}{2\chi_0} \vec{L}^2 \right],
\end{align}
where $\rho_s = S^2(I_1 - 2I_2) a_0^{2-d}$ is the spin stiffness and $\chi_0 = [2S^2(I_1 + 2I_2)a_0^d]^{-1}$ is the uniform susceptibility.

3. \textit{Kondo Interaction:} Substituting the spin decomposition into $H_K$ yields:
\begin{align}
H_K &= J_{K} S \sum_{i,a} (\eta_i \vec{n}_i + 2a_0^d \vec{L}_i) \cdot \vec{s}_{c,a,i}.
\end{align}
The first term couples the electron spin density to the staggered field $\eta_i \vec{n}$. In momentum space, this interaction scatters electrons from wavevector $\vec{k}$ to $\vec{k}+\vec{Q}$. We assume that the size of the Fermi surface is small compared to the antiferromagnetic wave vector $\vec{Q}$ (i.e., $2k_F < |\vec{Q}|$). Under this condition, low-energy electrons near the Fermi surface cannot accommodate the large momentum transfer required to scatter off the staggered field. Consequently, the spatial Kondo coupling is energetically suppressed. We neglect it and retain only the coupling to the uniform component:
\begin{align}
H_K \approx J_K \int d^dx \, \vec{L} \cdot \vec{s}_{c}.
\end{align}
where $\vec{s}_{c}=\sum_{a}\vec{s}_{c,a}$.

\vspace{0.5em}
\noindent\textbf{Effective Field Theory.}
Collecting all terms involving $\vec{L}$, we have a Gaussian integral:
\begin{align}
S[\vec{L}] = \int d\tau d^dx \left[ \frac{1}{2\chi_0} \vec{L}^2 + \vec{L} \cdot \left( -i \frac{S}{a_0^d} \vec{n} \times \partial_\tau \vec{n} + J_K \vec{s}_c \right) \right].
\end{align}
Integrating out $\vec{L}$ leads to:
\begin{align}
S_{\text{eff}} &= -\frac{\chi_0}{2} \int d\tau d^dx \left(- i \frac{S}{a_0^d} \vec{n} \times \partial_\tau \vec{n} + J_K \vec{s}_{c} \right)^2 \nonumber\\
&= \int d\tau d^dx \left[ \frac{\chi}{2} (\partial_\tau \vec{n})^2 + i \lambda_{K} (\vec{n} \times \partial_\tau \vec{n}) \cdot \vec{s}_c - \frac{\chi_0 J_K^2}{2} (\vec{s}_{c} \cdot \vec{s}_{c}) \right],
\end{align}
where we used $(\vec{n} \times \partial_\tau \vec{n})^2 = (\partial_\tau \vec{n})^2$. The effective susceptibility is $\chi \equiv \chi_0 S^2 / a_0^{2d}$, and the spin-motive force coupling is $\lambda_{K} \equiv \chi_0 S J_K / a_0^d$.

Combining this with the stiffness term $\frac{\rho_s}{2}(\nabla \vec{n})^2$, we obtain the total effective action. To match standard textbook notation, we define the spin-wave velocity $c$ and the coupling constant $g$ via:
\[
c = \sqrt{\frac{\rho_s}{\chi}}, \qquad \frac{1}{g} =  \chi.
\]
The final term $S_{\text{int}} \propto -(\psi^\dagger \vec{\tau} \psi)^2$ represents a four-fermion interaction.
Also using the spinor identity
$\vec{\tau}_{\alpha\beta} \cdot \vec{\tau}_{\gamma\delta} = 2\delta_{\alpha\delta}\delta_{\beta\gamma} - \delta_{\alpha\beta}\delta_{\gamma\delta}$,
the spin-spin interaction can be rewritten as a density-density interaction, qualitatively similar to a Hubbard $U$ term. Since the Fermi liquid interaction behaves qualitatively the same as a Fermi gas (merely renormalizing Landau parameters without changing the universality class), this term is non-singular for the critical dynamics and is therefore neglected. The final action takes the form:
\begin{align}
S &= S_{c} + \frac{1}{2g} \int_{x,\tau} \left[ c^2 (\nabla \vec{n})^{2} +  (\partial_{\tau}\vec{n})^{2} \right] 
+ i \lambda_{K} \int_{x,\tau} (\vec{n}\times \partial_{\tau}\vec{n})\cdot \vec{s}_{c} 
.
\end{align}
where $S_{c}$ is the action for the conduction electrons, as defined in the main text. In the following sections, we omit the channel index for simplicity, since the channel degrees of freedom do not affect the evaluation of the one-loop fermion-boson diagrams.

\subsection{Long-wavelength expansion}

To perform the renormalization group analysis in the dimensional regularization scheme, we must treat the non-linear constraint $|\vec{n}|^2=1$ perturbatively. We resolve the constraint by parameterizing the Néel field $\vec{n}$ in terms of $d-1$ massless Goldstone modes. We introduce the fluctuating Bose field $\vec{\pi}(\vec{x}, \tau)$ such that:
\begin{equation}
\vec{n}=\left(\vec{\pi},\sigma\right), \qquad \sigma=\sqrt{1-\vec{\pi}^2}= 1-\frac{1}{2}\vec{\pi}^2 - \frac{1}{8}(\vec{\pi}^2)^2 + \dots
\end{equation}
The Berry phase term governing the coupling to the itinerant electrons involves the cross product $\vec{n}\times \partial_{\tau}\vec{n}$. Using the parametrization above, this expands to:
\begin{equation}\label{kondoexp}
\vec{n}\times \partial_{\tau}\vec{n}=\begin{pmatrix}
\frac{1}{\sigma}\left(-\dot{\pi}_2-\dot{\pi}_1\pi_1\pi_2+\pi_1\pi_1\dot{\pi}_2\right)\\
\frac{1}{\sigma}\left(\dot{\pi}_1+\dot{\pi}_2\pi_2\pi_1-\pi_2\pi_2\dot{\pi}_1\right)\\
\dot{\pi}_2\pi_1-\pi_2\dot{\pi}_1\\
\end{pmatrix}
\cong 
\begin{pmatrix}
-\dot{\pi}_2 - \frac{1}{2}\pi^2\dot{\pi}_2 + \pi_2(\vec{\pi}\cdot\dot{\vec{\pi}}) \\
\dot{\pi}_1 + \frac{1}{2}\pi^2\dot{\pi}_1 - \pi_1(\vec{\pi}\cdot\dot{\vec{\pi}}) \\
\pi_1\dot{\pi}_2 - \pi_2\dot{\pi}_1
\end{pmatrix} + O(\pi^4).
\end{equation}

To organize the perturbative expansion, we rescale the field by the coupling constant $g$. We define the renormalized field $\vec{\phi}$ such that $\vec{\pi} = \sqrt{g}\vec{\phi}$. Under this rescaling, the kinetic term of the QNLSM becomes:
\begin{equation}
\frac{1}{2g}(\partial_\mu \vec{n})^2 = \frac{1}{2}(\partial_\mu \vec{\phi})^2 + \frac{g}{2}(\vec{\phi}\cdot\partial_\mu\vec{\phi})^2 + O(g^2).
\end{equation}
Similarly, substituting $\vec{\pi} = \sqrt{g}\vec{\phi}$ into Eq.~(\ref{kondoexp}) and coupling it to the electron spin density $\vec{s}_{c} = \psi^\dagger \vec{\sigma} \psi$, we generate the interaction vertices. The leading term (order $\sqrt{g}$) couples the time-derivative of the boson to the transverse electron spin, while the next term (order $g$) couples to the longitudinal spin:
\begin{align}
S_{int} &\sim i \int d\tau d^dx \, (\vec{n}\times\dot{\vec{n}}) \cdot \vec{s}_{c} \nonumber\\
&\sim i\sqrt{g} \int (\dot{\phi}_1 \sigma_y - \dot{\phi}_2 \sigma_x) + ig \int (\phi_1\dot{\phi}_2 - \phi_2\dot{\phi}_1)\sigma_z + \dots
\end{align}

\section{Renormalization group scheme}\label{sec3}
\subsection{Tree level scaling}
In this section we carry out the renormalization group(RG) analysis of the action (\ref{model}) by $\epsilon$ expansion, where $\epsilon=d-1$ and $d$ is the spatial dimensions. Our analysis is based on the combination of Wilson's bosonic RG and Shankar's fermionic RG\cite{Shankar}, which was shown to be valid when momentum and frequency scales the same for both bosons and fermions ($z=1$) \cite{Yamamoto-PRB}.  We first perform the tree level analysis, which can be done by simple dimensional counting. 

For the bosonic degree of freedom with the propagator, if we count the dimension by assigning $\left[p\right]=1 $, $\left[\omega\right]=1$ , then the scaling dimension of the bosonic field $\vec{\phi}$ is: 

\begin{equation}\label{scalingpi}
\begin{aligned}
&\left[\vec{\phi}\right]=-\frac{d+3}{2}
\end{aligned}
\end{equation}
which means after the rescaling $q\rightarrow bq$, $\omega\rightarrow b\omega$ , the  bosonic field rescales as $\vec{\phi}\rightarrow b^{-\frac{d+3}{2}}\vec{\phi}$ in order to make the the propagators invariant . By the scaling dimension (\ref{scalingpi}), one can easily check that:

\begin{equation}
\begin{aligned}
&\left[g\right]=-\epsilon=-\left(d-1\right)\\
\end{aligned}
\end{equation}

For the fermionic part, it has been recognized that the presence of the Fermi surface can influence the RG analysis in a dramatic manner \cite{Shankar}. We firstly linearize the dispersion $E_{\vec{K}}=v_Fk$, where $k=\vert\vec{K}\vert-K_F$ is the momentum relative to the Fermi momentum $K_F$, and $v_F$ is Fermi velocity.  The fermionic propagator is thus:

\begin{equation}
G_{\psi}\left(\vec{K},i\Omega\right)=\frac{1}{i\Omega-v_Fk}.
\end{equation}

Since the fermionic excitations lie within the vicinity of Fermi surface, after integrating out the fast modes whose momentum lying within $\left[\frac{\Lambda}{s}, \Lambda\right]$, where $\Lambda$ is the momentum cut-off, the integral $ \int^{\Lambda}d^dK$ becomes:

\begin{equation}
 \int^{\Lambda/s}d^dK\equiv \int d^{d-1}\Omega_{\vec{K}}\int^{K_F+\frac{\Lambda}{s}}_{K_F-\frac{\Lambda}{s}} K^{d-1}dK
 \end{equation} 
however, one can easily see that no simple rescaling of $K$ can restore the integral back to the original form. This difficulty results from the constraint that the momentums of the original fermionic theory is defined within the vicinity shell of the Fermi surface, so that simple rescaling of $K$ would break this momentum constraint.

To overcome this difficulty, it was pointed out that only the momentum perpendicular to Fermi surface should be rescaled\cite{Shankar}, since

\begin{equation}
\begin{aligned}
& \int^{\Lambda/s}d^DK\equiv \int d^{D-1}\Omega_{\vec{K}}\int^{K_F+\frac{\Lambda}{s}}_{K_F-\frac{\Lambda}{s}} K^{D-1}dK\\
&=\int d^{D-1}\Omega_{\vec{K}}\int^{\frac{\Lambda}{s}}_{-\frac{\Lambda}{s}} \left(k+K_F\right)^{D-1}dk\\
&\cong K_F^{D-1}\int d^{D-1}\Omega_{\vec{K}}\int^{\frac{\Lambda}{s}}_{-\frac{\Lambda}{s}}dk
&\end{aligned}
 \end{equation}

 Therefore, we have $\left[d^dK\right]=\left[dk\right]=1$ , and thus the scaling dimension of the fermionic field $\psi$ is:

\begin{equation}
\begin{aligned}
&\left[\psi\right]=-\frac{3}{2}\\
\end{aligned}
\end{equation}

where we count $\left[k\right]=1$, $\left[\Omega\right]=1$. Since the Landau damping term from fermion polarization is subleading compared with the bare boson dynamics, the boson and fermion both have dynamical exponent $z=1$. Therefore one can combine boson and fermion scaling together\cite{Yamamoto-PRB} to analyze the scaling dimension of the Kondo coupling.

The scaling dimension of the effective Kondo coupling is $[\lambda]=[\sqrt{g}\lambda_{K}]=-(d-1)/2$.

\subsection{Dimensional Regularization}
We perform the RG calculation in the dimensional regularization scheme\cite{fradkin2021}, using $\epsilon=d-1$ as an expansion parameter. 
By expressing the bare fields and couplings in Eq.3 of the main text in terms of the counter terms and renormalized couplings we obtain, 
\begin{align}
    \vec{\phi}&\rightarrow \sqrt{Z_{\phi}}\vec{\phi}, \;\psi\rightarrow \sqrt{Z_{\psi}}\psi, \nonumber\\  g&\rightarrow\mu^{-\epsilon}Z_{g}g, \;\lambda\rightarrow\mu^{-\frac{\epsilon}{2}}Z_{\lambda}\lambda
\end{align}
the action we start becomes 
\begin{align}
    S=&\int d\tau d^{d}x\{Z_{\psi}\psi^{\dagger}_{\sigma}(\partial_{\tau}+Z_{v}E(-i\vec{\nabla}))\psi_{\sigma}\nonumber\\
    &+\frac{1}{2}Z_{\phi}[(\partial_{\tau}\vec{\phi})^{2}+Z_{c}c^{2}(\vec{\nabla}\vec{\phi})^{2}]\nonumber
    \\
    &
    +\frac{1}{2}\mu^{-\epsilon}Z_{\phi}^2Z_{g}g[(\vec{\phi}\cdot\partial_{\tau}\vec{\phi})^{2}+c^{2}(\vec{\phi}\cdot\vec{\nabla}\vec{\phi})^{2}]\nonumber
    \\
    &+\mu^{-\frac{\epsilon}{2}}\sqrt{Z_{\phi}}Z_{\psi}Z_{\lambda}i\lambda\epsilon_{\alpha\beta}\partial_{\tau}\phi_{\alpha}\psi^{\dagger}\sigma^{\beta}\psi\nonumber\\
    &+\mu^{-\epsilon}Z_{\phi}Z_{\psi}Z_{\lambda_{z}}i\lambda_{z}\epsilon_{\alpha\beta}\phi_{\alpha}\partial_{\tau}\phi_{\beta}\psi^{\dagger}\sigma^{z}\psi\nonumber\\&+\mu^{-\frac{3\epsilon}{2}}Z_{\phi}^{\frac{3}{2}}Z_{\psi}Z_{\lambda'}i\lambda'(\phi_{\alpha}\partial_{\tau}\phi_{\alpha}\phi_{\beta}-\frac{1}{2}\vec{\phi}^2\partial_{\tau}\phi^{\beta})\epsilon_{\beta\gamma}\psi^{\dagger}\sigma^{\gamma}\psi+\cdots\label{dmaction}
\end{align}
where the dotted line contains higher order expansions. $\lambda_{z}=\sqrt{g}\lambda$,\,$\lambda'=g\lambda$ due to rotational symmetry.

To deriving the RG equations it is convenient to apply an infinitesimal staggered magnetic field term $S_h=-\frac{h}{g}\int d\tau d^dx\sigma=-\frac{h}{g}\int d\tau d^dx\sqrt{1-\vec{\pi}^2}=\frac{h}{g}\int d\tau d^dx\left(-1+\frac{1}{2}\vec{\pi}^2+\frac{1}{8}(\vec{\pi}^{2})^{2}\cdots\right)$ to the QNL$\sigma$M \cite{pelcovis1977,fradkin2021}. In the $\phi$ basis, they become \begin{align}
    S_{h}=Z_{\phi}Z_{h}\frac{h}{2}\int d\tau d^{d}x\vec{\phi}^2+Z_{\phi}^2Z_{h}Z_{g}\frac{hg}{8}\int d\tau d^d{x} (\vec{\phi}^2)^2.
\end{align}
where $Z_{h}=\sqrt{Z_{g}/Z_{\phi}}$ due to rotational symmetry.
The validity of this method relies on the fact that the QNLSM the action preserves the $O\left(3\right)$ symmetry.  
Even with the presence of non-vanishing Kondo coupling $\lambda_K$, one can still use this method to derive the coupling constant $g$, since under the $O\left(3\right)$ rotation, $\vec{n}\rightarrow R\vec{n}$, and thus $\vec{n}\times i\partial_{\tau}\vec{n}\rightarrow  R\left(\vec{n}\times i\partial_{\tau}\vec{n}\right)$, which can be compensated by a suitable $SU\left(2\right)$ transformation of the fermionic field $\psi\rightarrow U\psi$ where $U \in SU(2)$. 
As the result, the action (\ref{model})  still preserves the $O\left(3\right)$ symmetry even with Kondo coupling $\lambda_K$.

\subsection{One-loop RG equations}\label{sec4a}
We reserve the detailed derivation of the self-energies for Section III. Here, we summarize the resulting one-loop renormalization constants and vertex corrections.

The wavefunction renormalization constants ($Z_{\phi}, Z_{\psi}$) and the coupling renormalization constant ($Z_g$) are determined to be:
\begin{align}
    Z_{\phi} &=Z_{g}= 1+\frac{g}{2\pi c\epsilon}, \label{wfphi} \\
    Z_{\psi} &= 1+\frac{\lambda^{2}}{\pi\epsilon}\frac{c}{(v+c)^{2}},\label{wfpsi} \\
    Z_{c} &= Z_{v}=1. 
\end{align}
The one-loop vertex functions, $\Gamma_{\lambda}(\omega,q)$ and $\Gamma_{\lambda_z}(\omega,q)$, are calculated as:
\begin{align}
    \Gamma_{\lambda}(\omega,q) &= \omega\lambda\left\{1-\frac{\lambda_{z}}{\pi\epsilon}\left[\frac{c}{2(c+v)^2}+\frac{1}{c+v}\right]\right\} - ivq_{\perp}\frac{\lambda\lambda_{z}}{2\pi}\frac{c}{(c+v)^2}, \\
    \Gamma_{\lambda_z}(\omega,q) &= \omega\lambda_{z}\left\{1-\frac{g}{2\pi c\epsilon}-\frac{\sqrt{g}\lambda}{\epsilon}\left[\frac{c}{2\pi(c+v)^2}+\frac{1}{c+v}\right]\right\} - ivq_{\perp}\frac{\lambda_{z}^2}{2\pi}\frac{c}{(c+v)^2}.
\end{align}
These vertex functions are defined as the coefficient functions within the renormalized Kondo action $S_{K}$:
\begin{align}
    S_{K} \supset \int_{\omega,q} \Gamma_{\lambda}(\omega,q)\epsilon_{\alpha\beta}\phi^{\alpha}\psi^{\dagger}\sigma^{\beta}\psi + \int_{\omega,q} \epsilon_{\alpha\beta}\phi^{\alpha}\Gamma_{\lambda_{z}}(\omega_2,q_2)\phi^{\beta}\psi^{\dagger}\sigma^{z}\psi.
\end{align}

We observe that the vertex corrections generate terms proportional to $q_{\perp}$, representing momentum transfer normal to the Fermi surface. In real space with $d>1$, these terms correspond to highly non-local interactions rather than simple local derivative couplings. Since the original action (Eq.~3) defines a local field theory, these contributions would destroy the renormalizability of the model. To consistently truncate the theory and retain the local form of the action, we take the limit $v/c \rightarrow 0$, which suppresses these non-local terms.

\begin{figure}[!t]
\centering
\includegraphics[scale=.6]{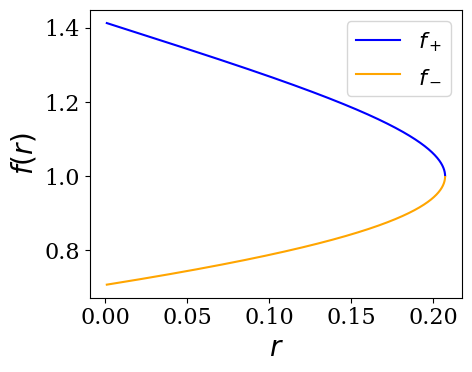}
\caption{Functions $f_{\pm}(r)$ determining the fixed points $\lambda^* = \sqrt{\epsilon}f_{\pm}(r)$ as a function of velocity ratio $r = v/c$. The upper branch $f_{+}$ corresponds to the Kondo-destruction QCP, while the lower branch $f_{-}$ represents the multi-critical point. The two solutions merge and vanish at $r=(\sqrt{2}-1)/2$.}
\label{fig:fpm}
\end{figure}

Using these counter-terms and requiring that the bare couplings remain independent of the renormalization scale $\mu$, we derive the beta functions for $g$, $\lambda$, and $\lambda_{z}$. The beta functions are defined with respect to the logarithmic scale $\ell \equiv -\ln\mu$:
\begin{align}
    \beta(g) \equiv \frac{dg}{d\ell} = -\mu\frac{dg}{d\mu}\bigg|_{\text{bare}}, \quad
    \beta(\lambda) \equiv \frac{d\lambda}{d\ell} = -\mu\frac{d\lambda}{d\mu}\bigg|_{\text{bare}}, \quad
    \beta(\lambda_{z}) \equiv \frac{d\lambda_z}{d\ell} = -\mu\frac{d\lambda_z}{d\mu}\bigg|_{\text{bare}}.
\end{align}
Thus, we obtain 
\begin{align}
    \beta(g)=&-\epsilon g+\frac{g^{2}}{2\pi c}\\
    \beta(\lambda)=&-\frac{1}{2}(\epsilon+\frac{g}{2\pi c})\lambda-2A\lambda^{3}+(A+2B)\sqrt{g}\lambda^{2}\\
    \beta(\lambda_z)=&\:\beta(\sqrt{g}\lambda).
\end{align}
where $A=\frac{c}{2\pi (c+v)^{2}}, B=\frac{1}{2\pi(c+v)}$. 
Rescaling $g\rightarrow 2\pi cg, \lambda\rightarrow \lambda/\sqrt{2A}$, we have
\begin{align}
    \beta(g)=&-\epsilon g+g^{2}\\
    \beta(\lambda)=&-\frac{1}{2}(\epsilon+g)\lambda-\lambda^{3}+\frac{3\sqrt{2}}{2}\frac{1+2r/3}{1+r}\sqrt{g}\lambda^{2}
\end{align}
where $r=v/c$ denotes for the ratio of Fermi velocity and spin velocity. The solutions of beta functions lead to the following fixed points:

\begin{itemize}\item \textit{Fixed Point I (Stable Antiferromagnet):}\begin{align}(g^*, \lambda^*) = (0, 0)\end{align}Describes the ordered antiferromagnetic phase (AF$_S$) where quantum fluctuations are irrelevant.\item \textit{Fixed Point II (Decoupled Magnetic QCP):}
\begin{align}
    (g^*, \lambda^*) = (\epsilon, 0)
\end{align}
Describes the conventional Wilson-Fisher magnetic transition between the antiferromagnet and a paramagnet with a small Fermi surface ($P_S$), decoupled from the Kondo effect.

\item \textit{Fixed Point III (Quantum Multicritical Point):}
\begin{align}
    (g^*, \lambda^*) = (\epsilon, \sqrt{\epsilon} f_{-}(r))
\end{align}
An unstable fixed point forming the nexus between the AF$_S$, $P_S$, and heavy-fermion ($P_L$) phases.

\item \textit{Fixed Point IV (Kondo-Destruction QCP):}
\begin{align}
    (g^*, \lambda^*) = (\epsilon, \sqrt{\epsilon} f_{+}(r))
\end{align}
The unstable critical point separating the magnetic order from the heavy Fermi liquid ($P_L$), characterizing the breakdown of Kondo screening at criticality.
\end{itemize}where $f_{\pm}(r)$ are given by:\begin{align}f_{\pm}(r)=\sqrt{\frac{1}{1+r}+\frac{1}{4(1+r)^{2}}
    \pm\sqrt{\left[\frac{1}{1+r}+\frac{1}{4(1+r)^{2}}\right]^{2}-1}}.\end{align}
 and are plotted in Fig.\ref{fig:fpm}

\subsection{Anomalous dimensions}

The anomalous dimensions for the boson and fermion fields are defined by the logarithmic derivatives of the field renormalization constants with respect to the energy scale $\mu$. Using the RG beta functions $\beta(g)$ and $\beta(\lambda)$, these can be expressed as:
\begin{align}
\gamma_{\phi} &= \frac{1}{2}\frac{d \ln Z_{\phi}}{d \ln\mu} = \frac{1}{2} \beta(g) \frac{\partial \ln Z_{\phi}}{\partial g}, \\
\gamma_{\psi} &= \frac{1}{2}\frac{d \ln Z_{\psi}}{d \ln\mu} = \frac{1}{2} \beta(\lambda) \frac{\partial \ln Z_{\psi}}{\partial \lambda}.
\end{align}
Evaluating these expressions at the RG fixed points yields the scale-invariant exponents:
\begin{align}
\gamma_{\phi} = \frac{g^*}{2}, \qquad \gamma_{\psi} = \frac{(\lambda^*)^2}{2}.
\end{align}
The specific values of these anomalous dimensions at the four distinct fixed points are discussed in the main text.
\section{One-loop diagrams}
\subsection{Classification of one-loop Feynman diagrams}
\begin{figure}[h!]
\centering
\includegraphics[width=0.55\textwidth]{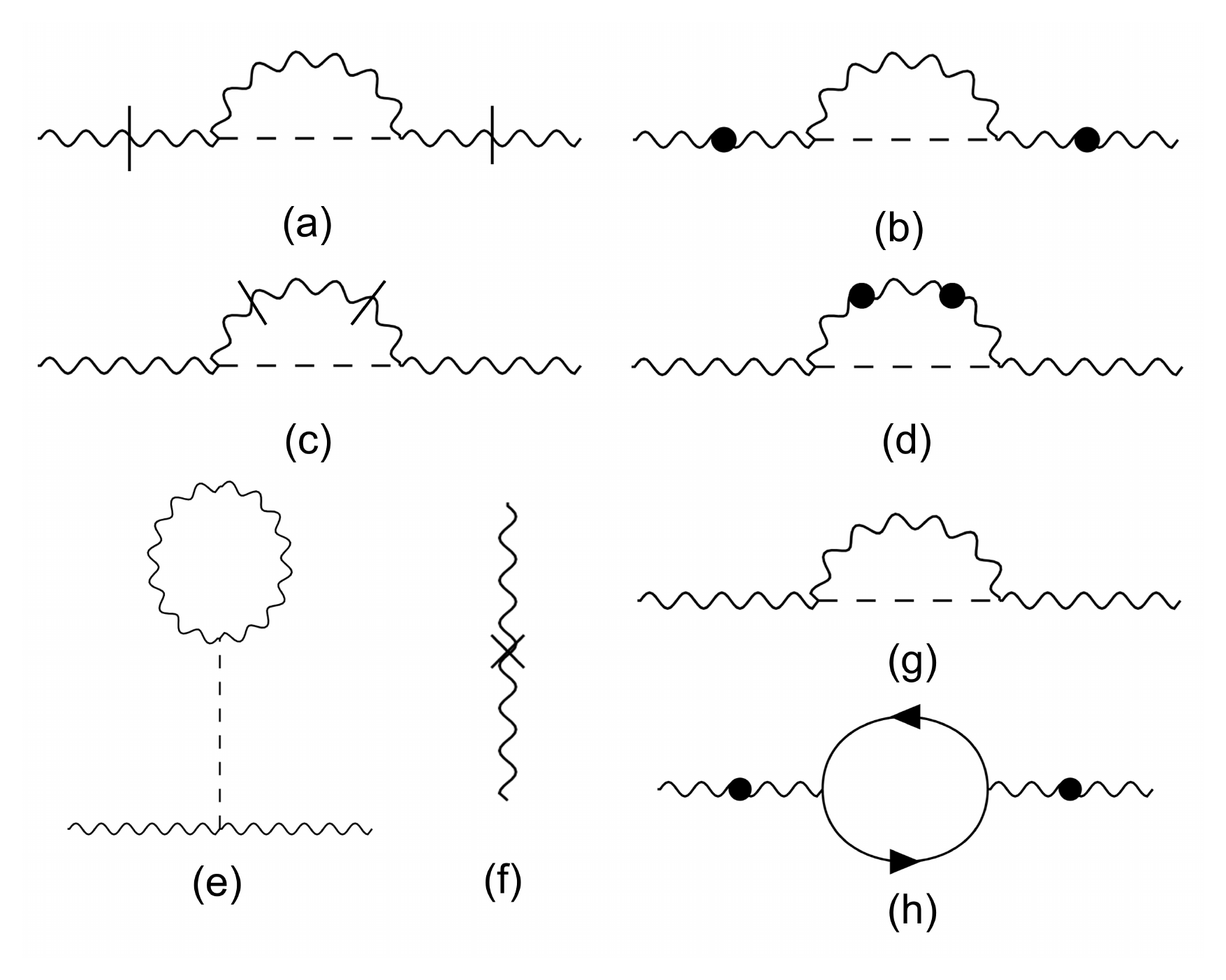}
\caption{One-loop Feynman diagrams contributing to the boson self-energy in the QNLSM with Kondo coupling.
(a)-(g) Purely bosonic corrections arising from the non-linear constraint and self-interactions.
Diagram (f) specifically represents the contribution from the path integral measure (Jacobian) obtained when relaxing the constraint $\vec{n}^2=1$.
The sum of diagrams (c), (d), (g), and (f) vanishes identically, ensuring the preservation of the underlying symmetries.
(h) The fermionic polarization bubble arising from the Berry phase contributed Kondo coupling $\lambda$, which generates weak Landau damping for the spin waves.}
\label{fig:bosonloop}
\end{figure}
\begin{figure}[h!]
\centering
\includegraphics[width=0.5\textwidth]{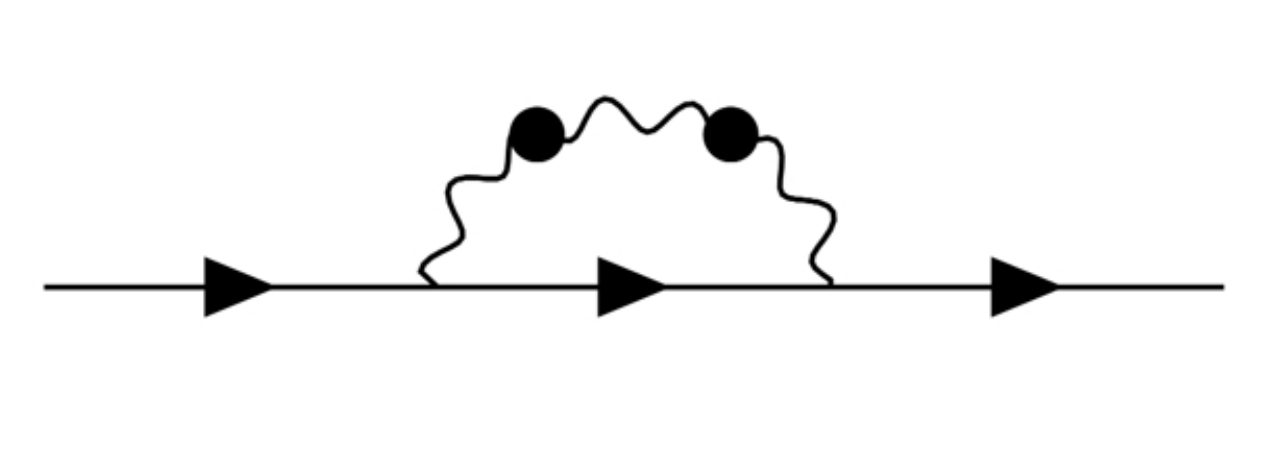}
\caption{One-loop fermion self-energy diagram.}
\label{fig:selfenergy}
\end{figure}

\begin{figure}[h!]
\centering
\includegraphics[width=0.55\textwidth]{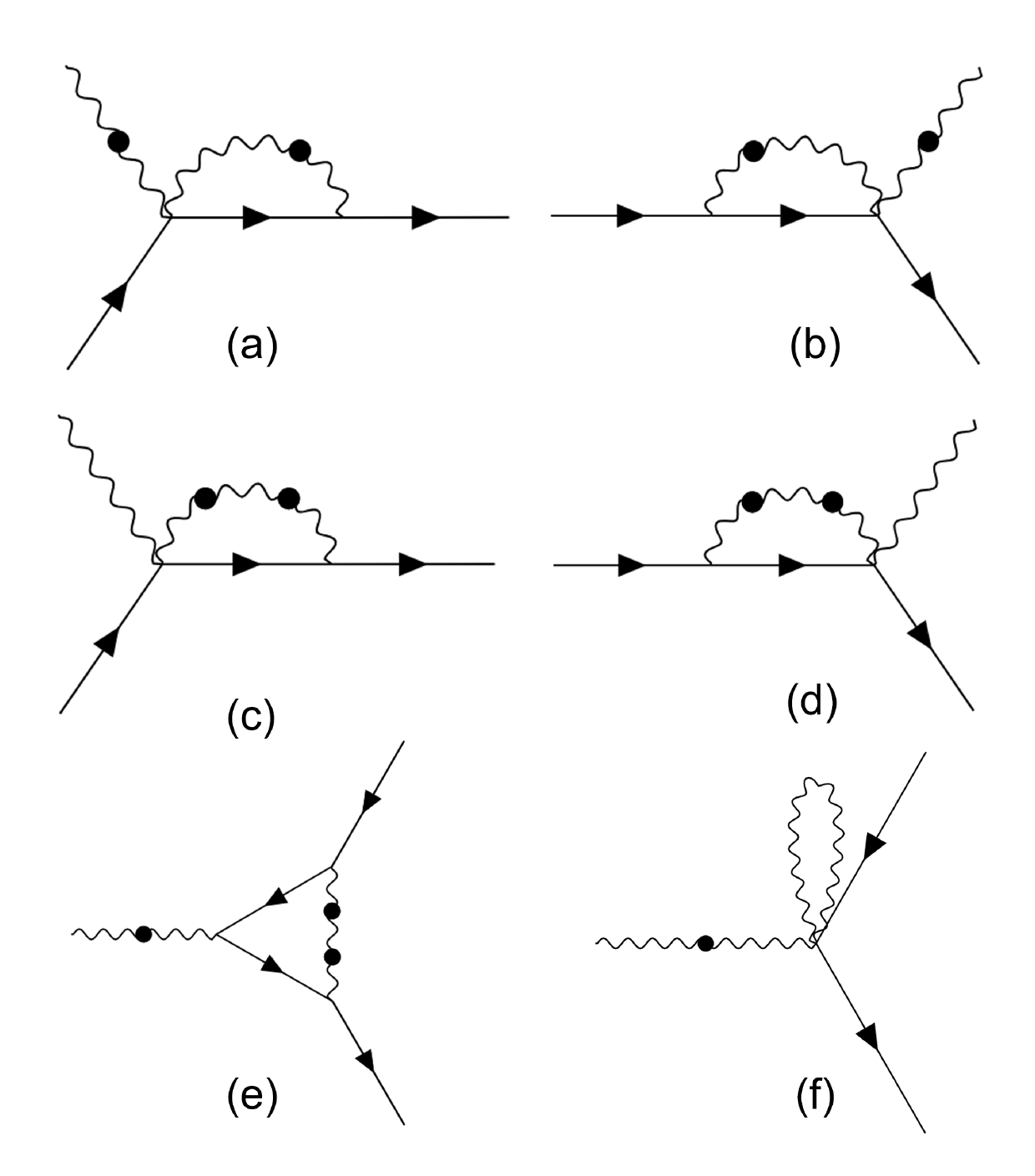}
\caption{Vertex corrections to the transverse Kondo coupling. Diagrams (a)–(d) contribute to the beta functions, whereas (e) and (f) vanish identically.}
\label{figs:transverse}
\end{figure}

\begin{figure}[h!]
\centering
\includegraphics[width=\textwidth]{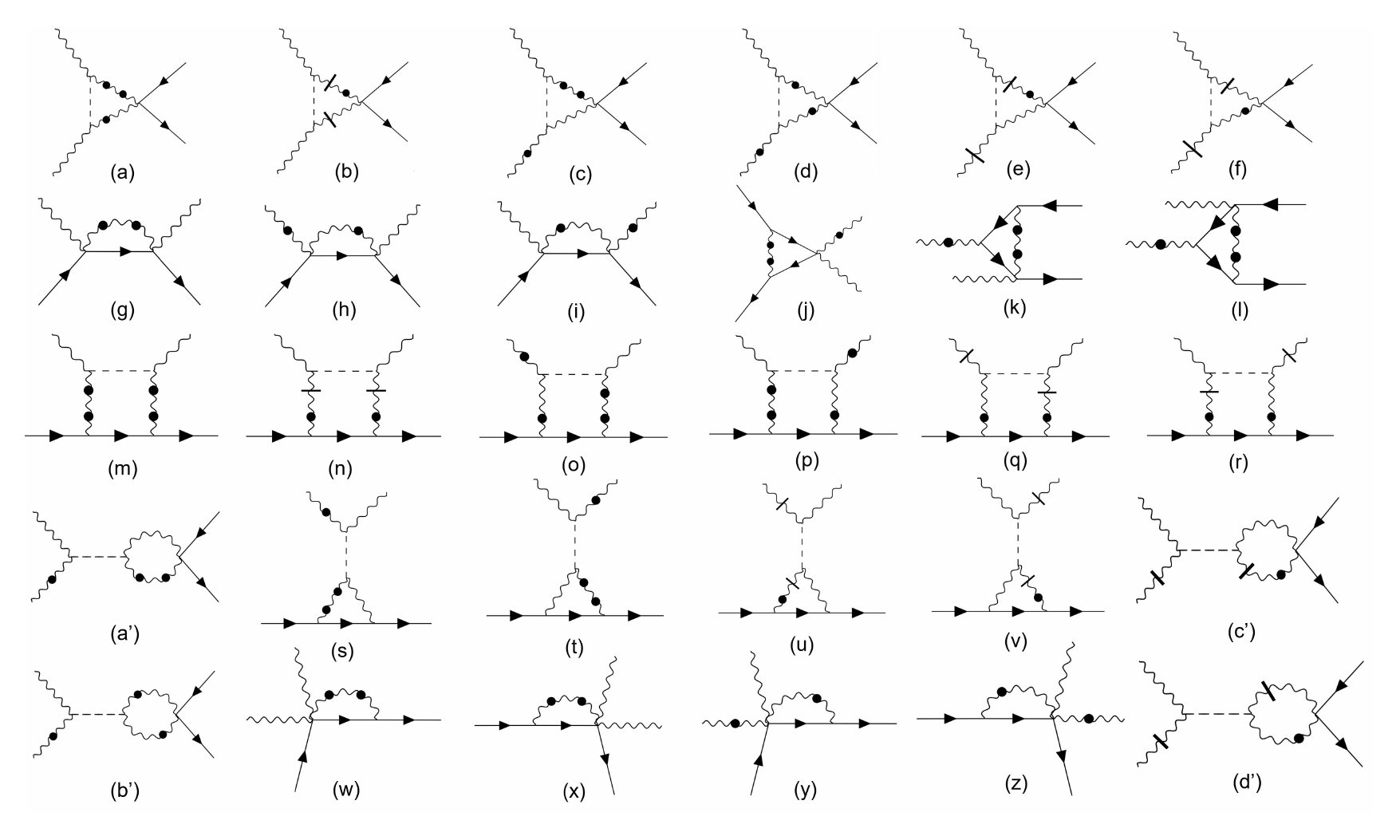}
\caption{
Full set of one-loop vertex corrections to the longitudinal Kondo coupling.  
Panels (a)–(z) and (a')-(d') represent the 30 distinct diagrams that contributes to the longitudinal corrections.  
Although these diagrams do not affect the RG flow of $\lambda$ at one loop,  
they collectively demonstrate that $\beta(\lambda_z)=\beta(\sqrt{g}\lambda)$,  
confirming that the symmetry relation $\lambda_z=\sqrt{g}\lambda$ is preserved under renormalization.}
\label{fig:longitudinal}
\end{figure}
In this subsection, we enumerate the one-loop Feynman diagrams and classify them based on their physical contributions: self-energy corrections to the tree-level propagators, vertex corrections to the transverse Kondo coupling $\lambda$, and vertex corrections to the longitudinal Kondo coupling $\lambda_z$.

Our general framework utilizes the cumulant expansion method to incorporate quantum fluctuations. The correction to the effective action, $\Delta S$, is given by:
\begin{align}
    \Delta S = \langle S_{\text{int}} \rangle_{c} - \frac{1}{2} \langle S_{\text{int}}^2 \rangle_{c} + \frac{1}{6} \langle S_{\text{int}}^3 \rangle_{c} + \cdots,
\end{align}
where $S_{\text{int}}$ represents the interacting part of the action and the subscript $c$ denotes the connected cumulant average.

Figure~\ref{fig:bosonloop} displays the diagrams responsible for the renormalization of the pure QNL$\sigma$M sector. Figure~\ref{fig:selfenergy} illustrates the fermion self-energy contributions. The detailed evaluation of these diagrams is provided in subsection B. Finally, Figure~\ref{figs:transverse} depicts the vertex corrections to the transverse Kondo coupling. In
Fig.\ref{figs:transverse}, 
\begin{align}
   \Delta S^{(a)}_{\lambda}+\Delta S^{(b)}_{\lambda}&=2i\lambda\lambda_{z}I'\int_{K,q} (-i\omega)\epsilon_{\alpha\beta}\phi^{\alpha}_{q}\psi^{\dagger}_{K+q,\mu}\sigma_{\mu\nu}^{\beta}\psi_{K,\nu}\\   
   &=-i\lambda\lambda_{z}\frac{1}{\pi(c+v)}\frac{1}{\epsilon}\int_{K,q} (-i\omega)\epsilon_{\alpha\beta}\phi^{\alpha}_{q}\psi^{\dagger}_{K+q,\mu}\sigma_{\mu\nu}^{\beta}\psi_{K,\nu}\\
   \Delta S^{(c)}_{\lambda}+\Delta S^{(d)}_{\lambda} &=-i\lambda\lambda_{z}\int_{K,q} [I(\vec{K}+\vec{q},\Omega+\omega)- I(\vec{K},\Omega)]\epsilon_{\alpha\beta}\phi^{\alpha}_{q}\psi^{\dagger}_{K+q,\mu}\sigma_{\mu\nu}^{\beta}\psi_{K,\nu}\\
   &=-i\lambda\lambda_{z}\frac{c}{2\pi(c+v)^2}\frac{1}{\epsilon}\int_{K,q} (-i\omega+vq_{\perp})\epsilon_{\alpha\beta}\phi^{\alpha}_{q}\psi^{\dagger}_{K+q,\mu}\sigma_{\mu\nu}^{\beta}\psi_{K,\nu}\\
   \Delta S^{(e)}_{\lambda}&=\Delta S^{(f)}_{\lambda}=0
\end{align}

The vertex corrections for longitudinal Kondo coupling are shown in Fig.\ref{fig:longitudinal}.
\begin{align}    \Delta S^{(a)}_{\lambda_{z}}&=\Delta S^{(b)}_{\lambda_{z}}=\Delta S^{(e)}_{\lambda_{z}}=\Delta S^{(f)}_{\lambda_{z}}=O(\omega^2,q^2),\\
\Delta S^{(a')}_{\lambda_{z}}&=\Delta S^{(b')}_{\lambda_{z}}=\Delta S^{(c')}_{\lambda_{z}}=\Delta S^{(d')}_{\lambda_{z}}=0,\\
\Delta S^{(c)}_{\lambda_{z}}&=\Delta S^{(d)}_{\lambda_{z}}=2ig\lambda_{z}J_{2}\int_{K,q_1,q_2}\epsilon_{\alpha\beta}\phi^{\alpha}_{q_1}(-i\omega_{2})\phi_{q_2}^{\beta}\psi^{\dagger}_{K+q_{1}+q_2}\sigma^{z}\psi_{K},\\
\Delta S^{(g)}_{\lambda_{z}}&=-2\lambda_{z}^2\int_{K,q_1,q_2}I(K+q_{1})\phi^{\alpha}_{q_1}\phi_{q_2}^{\beta}\psi^{\dagger}_{K+q_{1}+q_2}\psi_{K}, \, \Delta S^{(h)}_{\lambda_{z}}+\Delta S^{(i)}_{\lambda_{z}}=0,\\
\Delta S^{(j)}&+\Delta S^{(k)}_{\lambda_{z}}+\Delta S^{(l)}_{\lambda_{z}}=0,\\
\Delta S^{(m)}_{\lambda_{z}}&+\Delta S^{(n)}_{\lambda_{z}}=2g\lambda^2\int_{K,q_{1},q_{2}}I(K+q_{1})+O(q^2)\phi^{\alpha}_{q_1}\phi_{q_2}^{\alpha}\psi^{\dagger}_{K+q_{1}+q_2}\psi_{K},\\
\Delta S^{(o)}_{\lambda_{z}}&+\Delta S^{(p)}_{\lambda_{z}}=-2ig\lambda^2I_{1}\int_{K,q_1,q_2}\epsilon_{\alpha\beta}\phi^{\alpha}_{q_1}(-i\omega_{2})\phi_{q_2}^{\beta}\psi^{\dagger}_{K+q_{1}+q_2}\sigma^{z}\psi_{K},\\
\Delta S^{(q)}_{\lambda_{z}}&+\Delta S^{(r)}_{\lambda_{z}}=-2ig\lambda^2I_{2}\int_{K,q_1,q_2}\epsilon_{\alpha\beta}\phi^{\alpha}_{q_1}(-iq_{2\perp})\phi_{q_2}^{\beta}\psi^{\dagger}_{K+q_{1}+q_2}\sigma^{z}\psi_{K},\\
\Delta S^{(s)}_{\lambda_{z}}&+\Delta S^{(t)}_{\lambda_{z}}=\Delta S^{(u)}_{\lambda_{z}}+\Delta S^{(v)}_{\lambda_{z}}=0,\\
\Delta S^{(w)}_{\lambda_{z}}&+\Delta S^{(x)}_{\lambda_{z}}=0,\, \Delta S^{(y)}_{\lambda_{z}}+\Delta S^{(z)}_{\lambda_{z}}=4i\lambda'\lambda I'\int_{K,q_1,q_2}\epsilon_{\alpha\beta}\phi^{\alpha}_{q_1}(-i\omega_{2})\phi_{q_2}^{\beta}\psi^{\dagger}_{K+q_{1}+q_2}\sigma^{z}\psi_{K}
\end{align}

where \begin{align}
    J_{2}=&\int\frac{d^dpd\nu}{(2\pi)^{(d+1)}}\frac{\nu^2}{(\nu^2+c^2\vec{p}^2)^2}=\frac{1}{(d+1)c^d}\int\frac{d^dpd\nu}{(2\pi)^{(d+1)}}\frac{1}{(\nu^2+\vec{p}^2)}=-\frac{1}{4\pi c\epsilon}+\text{finite terms},\\
    I_{1}=&\int \frac{d\nu d^dp}{(2\pi)^{d+1}}\frac{i\nu^3}{(\nu^2+c^2\vec{p}^2)^2}\frac{1}{i\nu-vp_{\perp}}=-\frac{1}{2\pi\epsilon}\frac{1}{c+v}+\frac{1}{4\pi\epsilon}\frac{c}{(c+v)^2}+\text{finite terms},\\
    I_2=&\int \frac{d\nu d^dp}{(2\pi)^{d+1}}\frac{i\nu^2cp_{\perp}}{(\nu^2+c^2\vec{p}^2)^2}\frac{1}{i\nu-vp_{\perp}}=-i\frac{c}{v}(J_{2}-I_{1})=i\frac{v}{4\pi\epsilon}\frac{c}{(c+v)^2}+\text{finite terms}.\\
    I'=&\int \frac{d\nu d^{d}p}{(2\pi)^{d+1}}\frac{i\nu}{\nu^{2}+c^{2}\vec{p}^{\;2}}\frac{1}{i\nu-vp_{\perp}}=-\frac{1}{2\pi(v+c)}\frac{1}{\epsilon}+\text{finite terms}.
\end{align}
Collecting all the vertex corrections, the vertex function for longitudinal coupling is
\begin{align}
    \Gamma_{\lambda_{z}}(\vec{q},\omega)=&\omega(\lambda_{z}+2g\lambda_{z}J_{2}-2g\lambda^2I_1+4\lambda'\lambda I')-2g\lambda^2q_{\perp}I_2\\
    =&\omega\sqrt{g}\lambda\left\{1-\frac{g}{2\pi c\epsilon}-\frac{\sqrt{g}\lambda}{\pi\epsilon}\left[\frac{c}{2(c+v)^2}+\frac{1}{c+v}\right]\right\}-ivq_{\perp}\frac{g\lambda^2}{2\pi\epsilon}\frac{c}{(c+v)^2}.
\end{align}

\subsection{Diagrams that contribute to beta functions}
\subsubsection{Boson self-energy}

At the one-loop level, the inverse boson propagator takes the form:
\begin{align}
    D^{-1}(\vec{q},\omega) = D_{0}^{-1}(\vec{q},\omega) + \Pi(\vec{q},\omega),
\end{align}
where the bare propagator is defined as $D_{0}^{-1}(\vec{q},\omega) = h + \omega^2 + c^{2}\vec{q}^{\,2}$. The boson self-energy $\Pi(\vec{q},\omega)$ encompasses contributions from both self-interaction and coupling to the fermions:
\begin{align}
   \Pi(\vec{q},\omega) =& \, g\left(\omega^{2}+c^{2}\vec{q}^{\,2}+h\right)\int\frac{d\nu d^{d}p}{(2\pi)^{d+1}}D_0(\vec{p},\nu) \nonumber \\
   &- g\int \frac{d\nu d^{d}p}{(2\pi)^{d+1}} + g\int\frac{d\nu d^{d}p}{(2\pi)^{d+1}}(\nu^2+c^2p^2+h)D_0(\vec{p},\nu) \nonumber \\
   &+ M\lambda^{2}N_{F}\omega^{2}\left(1-\frac{|\omega|}{\sqrt{\omega^{2}+v_{F}^{2}\vec{q}^{\;2}}}\right).
\end{align}
The first term corresponds to the bosonic tadpole contributions (Diagrams (a), (b), (e) in Fig.~\ref{fig:bosonloop}). The terms in the second line originate from the measure and metric corrections (Diagrams (c), (d), (f), (g)); remarkably, these terms cancel each other exactly, preserving the non-linear constraint symmetries. The final term represents the fermion polarization bubble (Fig.~\ref{fig:bosonloop}(h)) in two spatial dimensions.

While the fermion loop generates Landau damping, it does not introduce new ultraviolet divergences to the boson stiffness $g$ at this order, and the correction is weak compared with bare propagators. In the framework of dimensional regularization with $d = 1+\epsilon$, the singular part of the self-energy arises solely from the bosonic sector, yielding a simple pole in $\epsilon$:
\begin{align}
    \Pi_{\text{sing}}(\vec{q},\omega) = -\frac{g}{2\pi\epsilon}(\omega^{2}+c^{2}\vec{q}^{\,2}+h).
\end{align}

Therefore, at one loop level, the boson self-energy is solely determined by the QNLSM side.

\subsubsection{Fermion self-energy}
The fermion self-energy at one loop level is determined by the transverse part of the Kondo action in Fig.\ref{fig:selfenergy},
\begin{align}
    &G^{-1}(\vec{K},\Omega)=i\Omega-E(\vec{K})-\Sigma(\vec{K},\Omega)\\
    &\Sigma(\vec{K},\Omega)\nonumber\\=&-2\lambda^{2}\mu^{-\epsilon}\int \frac{d\nu d^{d}p}{(2\pi)^{d+1}}\frac{\nu^{2}}{\nu^{2}+c^{2}\vec{p}^{\;2}}\frac{1}{i(\Omega+\nu)-E(\vec{K}+\vec{p})}
\end{align}
where $E(\vec{K}+\vec{p})= v(k_{\perp}+p_{\perp})+\frac{{\vec{p}}^{2}+k_{\perp}^{2}}{2m}$, and $\perp$ and $\parallel$ denote for directions perpendicular and parallel to the Fermi surface respectively, with $\vec{K}=\hat{n}(K_{F}+k_{\perp}),\vec{p}=\hat{n}p_{\perp}+\vec{p}_{\parallel}$.  Since $k_{\perp}$ and $\vec{p}_{\parallel}$ scale the same as $k_{\perp}$ and $p_{\perp}$, the curvature effects are irrelevant; thus, we approximate $E(\vec{K})= vk_{\perp}$ in the low energy limit.

Using dimensional regularization, we find that $\Sigma$ is also proportional to $1/\epsilon$,
\begin{align}
\Sigma(\vec{K},\Omega)=\frac{\lambda^{2}}{\pi}\frac{c}{(v+c)^{2}}\frac{i\Omega-
vk_{\perp}}{\epsilon}
\end{align}
Here we can see that only wave function renormalizations receive quantum corrections and the Fermi velocity is unchanged during RG.

\subsubsection{Vertex correction}
Due to the spin-spin interaction form of the Kondo coupling, the vertex correction from pure transverse part of the Kondo couplings vanishes (c.f. Fig.\ref{figs:transverse}  (e,f)). The transverse part $\lambda$ only receives vertex correction from combination of transverse and longitudinal vertices (c.f. Fig.\ref{figs:transverse}  (a-d)):
\begin{align}
    \Gamma_{\lambda}(\vec{q},\omega)
=&\omega\lambda\left\{1-\frac{\lambda_{z}}{\pi\epsilon}\left[\frac{c}{2(c+v)^2}+\frac{1}{c+v}\right]\right\}-ivq_{\perp}\frac{\lambda\lambda_{z}}{2\pi}\frac{c}{(c+v)^2}.
\end{align}
And the correction of longitudinal part of Kondo coupling is obtained by collecting diagrams in Fig.\ref{fig:longitudinal},
\begin{align}
    \Gamma_{\lambda_{z}}(\vec{q},\omega)
    =\omega\lambda_{z}\left\{1-\frac{g}{2\pi c\epsilon}-\frac{\sqrt{g}\lambda}{\pi\epsilon}\left[\frac{c}{2(c+v)^2}+\frac{1}{c+v}\right]\right\}  -ivq_{\perp}\frac{\lambda_{z}^2}{2\pi}\frac{c}{(c+v)^2}.
\end{align}

\subsubsection{Calculation of loop integrals}
We first list some useful integrals,
\begin{align}
    &\int\frac{d^{d}q}{(2\pi)^{d}}\frac{1}{m^{2}+c^{2}q^{2}}=\frac{m^{d-2}}{(4\pi)^{d/2}c^{d}}\Gamma(1-d/2)\label{eq1}\\
    &\int \frac{d\omega d^{d-1}q}{(2\pi)^{d}}\frac{\omega^2}{(A\omega^{2}+Bq^{2}+m^2)^n}=\frac{\pi^{\frac{d}{2}}\Gamma(n-1-\frac{d}{2})}{2\Gamma(n)}\frac{1}{\sqrt{A^{3}B^{d-1}}m^{2n-2-d}}\\
    &\frac{1}{A^{\alpha}B^{\beta}}=\frac{\Gamma(\alpha+\beta)}{\Gamma(\alpha)\Gamma(\beta)}\int_{0}^{1}du\frac{u^{\alpha-1}(1-u)^{\beta-1}}{[uA+(1-u)B]^{\alpha+\beta}}\label{eq2}
\end{align}
The fermion self energy could be written as
\begin{align}
&\Sigma(\vec{K},\Omega)\nonumber=-2\lambda^{2}\mu^{-\epsilon}I(\vec{K},\Omega)\\
&I(\vec{K},\Omega)=-\int \frac{d\omega d^{d}p}{(2\pi)^{d+1}}\frac{\omega^{2}}{\omega^{2}+c^{2}\vec{p}^{\;2}}\frac{i(\omega+\Omega)+v(p_{\perp}+k_{\perp})}{(\omega+\Omega)^{2}+v^{2}(p_{\perp}+k_{\perp})^{2}}.
\end{align}
Integrating over a $(d-1)$ dimensional hypersurface $\vec{p}_{\parallel}$ with the help of Eq.(\ref{eq1}) gives,
\begin{align}
    I=-\frac{\Gamma[(3-d)/3]}{(4\pi)^{(d-1)/2}c^{d-1}}\int\frac{d\omega d p_{\perp}}{(2\pi)^{2}}\frac{\omega^{2}}{(\omega^2+c^{2}p_{\perp}^{2})^{(3-d)/2}}\frac{i(\omega+\Omega)+v(p_{\perp}+k_{\perp})}{(\omega+\Omega)^{2}+v^{2}(p_{\perp}+k_{\perp})^{2}}
\end{align}
Next we utilize Feynman parametrization in \ref{eq2},
\begin{align}
    I=-\frac{\Gamma[(3-d)/3]}{(4\pi)^{(d-1)/2}c^{d-1}}\int_{0}^{1}du (1-u)^{(1-d)/2}\int\frac{d\omega d p_{\perp}}{(2\pi)^{2}}\frac{[i(\omega+\Omega)+v(p_{\perp}+k_{\perp})]\omega^{2}}{[\omega^{2}+(v^{2}u+c^2(1-u))p^{2}+u(2\omega\Omega+2v^{2}p_{\perp}k_{\perp}+\Omega^{2}+v^{2}k^{2})]^{(5-d)/2}}
\end{align}
Shifting the variables $\omega\rightarrow\omega-u\Omega, p_{\perp}\rightarrow p-\frac{uv^{2}k_{\perp}}{v^{2}u+c^{2}(1-u)}$ and integrate out $p_{\perp}$ and $\omega$, we get the following result near $d=1+\epsilon$,
\begin{align}
    I(\vec{K},\Omega)=-\frac{c}{2\pi(c+|v|)^{2}}\frac{i\Omega-vk_{\perp}}{\epsilon}+\text{finite terms}
\end{align}

Similarly,
\begin{align}
    I'=&\int \frac{d\omega d^{d}p}{(2\pi)^{d+1}}\frac{i\omega}{\omega^{2}+c^{2}\vec{p}^{\;2}}\frac{1}{i(\omega+\Omega)-v(p_{\perp}+k_{\perp})}\\
    =&\int \frac{d\omega d^{d}p}{(2\pi)^{d+1}}\frac{\omega^{2}}{\omega^{2}+c^{2}\vec{p}^{\;2}}\frac{1}{\omega^{2}+v^{2}p_{\perp}^{2}}=-\frac{1}{2\pi(v+c)}\frac{1}{\epsilon}+\text{finite terms}
\end{align}

\end{document}